\documentclass[12pt,twocolumn]{article}
\usepackage{titling}
\usepackage{authblk}
\usepackage{amssymb, amsmath}
\usepackage[caption=false]{subfig}
\usepackage{natbib}
\usepackage{graphicx} 
 \usepackage{floatpag}
 
\title{Toward a Zero-Parameter Model for Galaxy Rotation Curve Data}
\author[1]{S.\,~Cisneros}
\author[2]{J.\,~G.\,~O'Brien}
\author[1]{N.\,~S.~Oblath}
\author[1]{J.\,~A.~Formaggio}
\author[3]{M.\,Crowley}
\author[3]{K.\,Mikulski}
\affil[1]{Laboratory for Nuclear Science, Massachusetts Institute of Technology, Cambridge, MA~02139, USA}
\affil[2]{ Department of Sciences, Wentworth Institute of Technology,
Boston, MA~02130, USA}
\affil[3]{Department of Physics, University of Massachusetts Boston,
Boston, MA~02125, USA}
\date{}                     
\date{}
\begin{document}
\begin{titlingpage}
\maketitle

\begin{abstract}
Modeling   the luminous mass components of    spiral galaxies   in standard gravity  poses a challenge due to the missing mass problem.  However, with the addition of cold dark matter, the missing mass problem can be circumvented at the cost of additional free parameters to the theory.  The Luminous Convolution Model (LCM) reconsiders how we interpret rotation curve data, such that Doppler-shifted spectra measurements  can constrain  luminous mass discovery. For a sample of 25  galaxies of varying morphologies and sizes, we demonstrate an ansatz for  relative galaxy curvatures that  can  explain the missing mass.   We solve  for the  LCM free parameter, which we report    as  a ratio of radial densities of the emitter, to receiver galaxy baryonic mass, to an  exponent  of $1.63$. Here, we show that this exponent is sensitive to which Milky Way luminous mass model one chooses.  We then make a first prediction regarding the Milky Way mass profile in the inner one  kpc.
  Thus, with a bound on the LCM free parameter, we pave the way for future work, where the LCM will tested as a zero-parameter model  to predict   luminous mass from rotation curve data.  
\end{abstract}
\end{titlingpage}

\section{}
Light from  spiral galaxies  is considered by two different observational techniques. The first,   photometry,  is a measure of  total luminosity  in specific wavelength bands used to trace mass fractions (specific star populations, gas fractions,  dust, etc.).   From the   photometry,      mass is associated with 
total light by  population synthesis modeling (PSM).  PSM is  under-constrained \citep{conroy}   precisely  because the second observation of light,   
 Doppler-shifted spectra, yielded the dark matter problem. 
 
   \vspace{.5cm}
   
PSM returns  mass-to-light ratios (M/L) for galaxies, based on a   complex suite of model-based assumptions regarding chemical composition,   evolution, dust, gas fractions, metallicity,  etc.   Finding  a M/L  ratio  relies upon distance estimates, based on  underconstrained standard candles.    The M/L for a given galaxy 
 parametrizes  enclosed mass as a function of radius, which  then gives test particle motion by
  the Poisson equation.   Resulting 
   rotation curves are the expected Keplerian, declining  velocities $v_{lum}(r)$ seen in   Fig.~(\ref{galaxiesSmallest}).

  The conflicting observation, Doppler-shifted spectra of characteristic atomic transitions, is    interpreted in the Lorentz framework
   as  a  relative velocity between  frames.  The resulting   velocity parameter $v_{obs}$ can be seen to be essentially constant at large radii, beyond all  stars  (Fig.~\ref{galaxiesSmallest}). 
   The divergence of the velocities implied by shifted spectra and photometry
   gave rise to  the dark matter problem in spiral galaxies,  discovered by 
~\cite{1978Rubin} and ~\cite{Bosma78}. The mismatch between   dynamical mass  and the luminous mass is generally explained by a class of cold, dark matter halo models 
~\citep{NFW}. There have been successful attempts to test   alternate  theories such as Modified Newtonian Dynamics (MOND)~\citep{Milgrom} and Conformal Gravity \citep{fitting}.  
Since  dark matter halo models have  two free parameters,   they do little to constrain   population synthesis models at the present time \citep{conroy}, and hence serve as a possible interpretation,  not as a prediction.  
 
\vspace{.5cm}

Soon after the discovery of the flat-rotation curve problem, another curious trend was  noted by  \cite{Rub}.  In   \citet{salucci},  it was shown  with exquisite precision---in  a sample of  1,100 galaxies---that  rotation curves fall into a    spectrum   inflected     about  the Milky Way's rotation curve. 
      This spectrum 
  is known as the 
  Universal Rotation Curve  (URC);  it  is interpreted that galaxies smaller than the Milky Way are dark matter dominated, and those larger   require  only minimal dark matter halos.

   As dark matter particles are hypothesized to  interact in the  standard gravitational manner, there exists no physical reason for smaller galaxies to  accrete dark matter halos more successfully than larger  galaxies. 
   Likewise, the unexplained position of the Milky Way in the middle of the 
     URC spectrum is suggestive.   Thus,  we instead interpret the URC as indicative of frame-dependent effects due to our  observing frame,  the  Milky Way Galaxy,  in the    heuristic construction of the Luminous Convolution Model (LCM). This will serve as the starting point for the LCM interaction,  and we show   that the   URC spectrum can be explained   compactly   based on this model prescription.
  
  
The LCM   is  a fitting prescription,  with 
  one free parameter $b$, which we show  in this work can be reduced to a   constant dependent on the  specific choice of the
    Milky Way luminous mass model.    The  LCM ansatz is that  the relative,  very small gravitational curvatures of very large frames   are convolved in our observations of shifted spectra.   This statement is reminiscent of the MOND paradigm  \citep{Milgrom}, though it  exploits  changing relative curvatures (LCM) as opposed to modified accelerations (MOND).

\vspace{.5cm}

Usually, relative galaxy curvature effects are ignored in the flat-rotation curve problem, as   they are about four orders of magnitude smaller than the relative velocity effects when phrased kinematically.   Commonly,   relative curvature effects are evaluated (and obviated)  by taking the algebraic difference of the gravitational redshifts of a   galaxy from those of the Milky Way \citep{MTW}.   This is  a  Galilean transformation of frames, and   the    photon wave-vector should instead transform   within the  Lorentz group.   Hence, we instead  
 use  photon redshift frequencies to define  relative  galaxy frames, related through the    Lorentz  boost architecture to quantify the relationship kinematically.
   Kinematic phrasing of gravitational redshifts is common in astrophysics, what is new   is  the   treatment of  \emph{relative} curvatures.

  In this paper,  the  chosen  sample of 25 galaxies  is selected  in an effort to represent a full spectrum of galaxy sizes and morphologies, and no other bias was placed on selection.  
This sample was used to  both identify  the LCM free parameter and constrain the Milky Way luminous mass model.
We   propose that when the LCM  reproduces  
    rotation curve data perfectly,  as is the case in the current sample,  that 
    the resulting luminous mass profiles demonstrate  agreement between  the two observations of Dopper-shifted spectra and photometry. 
    Hence, the LCM introduces  a constraint to luminous mass modeling directly 
     from Doppler-shifted spectra, using luminous matter estimates within the  range defined by photometry.       By considering the small frame effects due to  relative curvatures of spiral galaxies, the LCM is a  successful fitting formula which   offers a 
   simple explanation for  the   URC spectrum    based   on 
 luminous mass estimates.

 This paper is organized as follows: Sec.~\ref{sec:DERIVE} sketches the heuristic construction of the LCM fitting formula, Sec.~\ref{fitting} describes the  fitting method, 
 Sec.~\ref{sample} offers analysis of the results, and Sec.~\ref{sec:conclusion} details conclusions and future directions. 
 
  \section{ LCM fitting formula }
\label{sec:DERIVE} 
 
 Dark matter (DM) theories interpret the discrepancy between rotation curve data  and Keplerian photometry predictions   as 
  the missing mass problem.    The   dark mass  is hypothesized to  interact  only gravitationally with ordinary baryonic matter.  
 This assumption is treated with the   rotation curve formula,
  \begin{equation}
v_{rot}^2 =  v_{lum}^2 +  v_{dark}^2,
\label{eq:zonte1}
\end{equation} 
where  the prediction $v_{rot}$ is fitted to reported rotation curve velocity data ($v_{obs}$ from shifted spectra  $\omega'$).      The term $v_{dark}$ is    the contribution to rotation due to a spherically symmetric   dark matter halo,   and $v_{lum}$ is from the baryonic contribution.   The rotation curve formula is  a mass sum $ M_{dynamical} = M_{lum}+M_{dark}$, as terms in $v^2$ represent  centripetal accelerations due to enclosed mass as a function of radius. The functional forms of the  terms in Eq.~(\ref{eq:zonte1})  have been well-established  (\cite{mfull}) and usually provide an accurate statistical fit to the data due to the presence of two free parameters per galaxy in $v_{dark}$.   

\subsection{LCM   construction}

 The LCM    modifies   Eq.~(\ref{eq:zonte1}),  by  
   replacing the 
DM halo velocity contributions, $v^2_{dark}$, with the relative curvature  convolution,  $\tilde{v}^2_{lcm} $, in the fitting formula: 
  \begin{equation}
v_{rot}^2 =  v_{lum}^2+ \alpha \tilde{v}_{lcm}^2 ,
\label{eq:zonteLCM}
\end{equation}  
 where $\alpha$ is a dimensionless  fitting parameter  and $\tilde{v}^2_{lcm} $is a function of luminous mass only.
 
We find in this work (see Fig.~\ref{fig:alphaDistrib1}) that the free parameter is    correlated with the dimensionless ratio,
\begin{equation}
\alpha=\left(\frac{\rho_{mw}}{\rho_{gal}}\right)^{b}.
\label{correl}
\end{equation}
The subscript $mw$ indicates the galaxy receiving the photon    and  the subscript $gal$ indicates the photon emitting  galaxy. 
  Radial densities $\rho_{mw}$ and $\rho_{gal}$ are    defined by: 
 \begin{equation}
 \rho_r=\frac{M_{total}}{r_e},
 \label{radDens}
 \end{equation}
where  the total  integrated   luminous mass at the limit of the photometric data is $M_{total}$,  and $r_e$ is the exponential galactic disk scale length from fitting the resulting LCM luminous mass profile with the standard form of the galactic thin disk.  Hence, for a given choice of the MW we can establish the free parameter space and fit for the exponent $b$, as shown in 
Fig.~\ref{fig:alphaDistrib1}.
 We report for a given Milky Way the exponent $b$ appears to be a constant.

  \vspace{.5cm}

 The  LCM  mapping  $\tilde{ v}_{lcm}^2$ is  a convolution of two Lorentz-type transforms $v_1$ and $v_2$:
\begin{equation}
\tilde{ v}_{lcm}^2=   \kappa^2  v_{1}  v_{2},
\label{eq:convolutionFunc}
\end{equation}
 scaled by  a   curvature ratio  $\kappa$,
\begin{equation}
\kappa(r)=\frac{c-\tilde{c}_{gal}(r)}{c-\tilde{c}_{mw}(r)}.
\label{eq:kappa}  
\end{equation}

Coordinate light speeds 
   $\tilde{c}_{gal} (r)$  and  $\tilde{c}_{mw} (r)$ are defined by 
the  pseudo-index of refraction $n(r)$  \citep{Narayan},  
    \begin{equation}
 n(r) \tilde{c }_{frame}=\left(\frac{1}{\sqrt{-g_{tt(r)}} }\right)_r  \tilde{c}_{frame}=c.
 \label{eq:index}
\end{equation}
and represent deviations from flatness for some mass distribution 
   which is symmetric about a central value of $r=0$.   The term  $g_{tt}$ is defined as the 
   Schwarzschild time metric coefficient.

The term $v_1$ is the relative galaxy-to-galaxy  transform.   It is based on 
comparing  the    standard  frequency form of the Lorentz Doppler shift formula  to its 
hyperbolic form,    
  \begin{equation}
 \frac{v_{obs}(r)}{c}=
\frac{  \frac{\omega'(r)}{\omega_o}- \frac{\omega_o}{\omega'(r)}}{ 
\frac{\omega'(r)}{\omega_o}+ \frac{\omega_o}{\omega'(r)}}\\
=\frac{e^\xi - e^{-\xi}}{ e^\xi + e^{-\xi}}
\label{eq:dataLorentz}
\end{equation}
and  identifying 
  the Lorentz boost exponential $e^\xi $ with the frequencies sent   and received, respectively 
  $ \omega_o$   and $\omega'$.

   We define slightly curved galaxy manifolds by the  Schwarzschild gravitational redshift 
     frequencies  $\omega (r)$ 
  \begin{equation}
 \frac{\omega_o }{ \omega ( r) }=\left(\frac{1}{\sqrt{-g_{tt}} }\right)_r,
\label{eq:Clone}
\end{equation}
where $\omega_o$  is  the   characteristic photon frequency.   

We use the  the weak field form of the time metric coefficient $g_{tt}$~\citep{Hartle}   
 \begin{equation}
g_{tt}(r) \approx -1 + 2\frac{\Phi(r)}{c^2},
\label{eq:weakfield}
\end{equation}
where    $\Phi$ is the  Newtonian  scalar   gravitational potential defined by the luminous mass reported from photometry.

  In Special Relativity,  the two frames involved in Lorentz transformation are perfectly  symmetric, so there  is no distinction between who is the emitter  and who  is the receiver of the photon.  However,  as we transition to 
    slightly curved galaxy frames,
  we pin the   Lorentz frames with 
redshift frequencies emitted and received  
 \begin{equation}
  e^\xi= \frac{\omega'_{receiver}}{ \omega_{emitter}}, 
  \label{XO}
  \end{equation}
for consistency.    It is common practice to use the functional form in Eq.~\ref{eq:dataLorentz} to rephrase gravitational redshifts kinematically  ~(\cite{Cisn,Radosz}). 

So, the term  $v_1$ is based on   the Lorentz exponential term: 
   \begin{equation}
e^{\xi_{c}}(r)=\frac{\omega_{mw}(r)}{\omega_{gal}(r)}.
\label{eq:11}
\end{equation}
  
Here  $\omega_{mw}$ and $\omega_{gal}$,  are  compared one-to-one in radii, and are 
the gravitational redshifts for $mw$   the receiving galaxy  (MW)   and   $gal$   the   photon emitting galaxy.    It is this pseudo-rapidity angle  $\xi_c$ which explains trends in the Universal Rotation Curve of  ~\cite{salucci}. 

  The  transformation  $v_1$ is  
      \begin{equation}
\frac{v_{1}}{c}=\left(\frac{1}{\cosh \xi_{c} } -1\right)=\left(\frac{ 2}{e^{\xi_{c}}+ e^{-\xi_{c}}} -1\right). 
\end{equation}
which can be interpreted as the difference from 
unity of the  
coordinate time of the 
  galaxy emitting the photons with respect to the 
  proper time of the 
  Milky Way.

The second  transformation, $v_{2}$,  is a   Lorentz boost which  transforms between  the curved $2$-frame $\xi_c$ (Eq.~\ref{eq:11})   to  the 
 flat $2$-frame  where observations are made.
 By the  constancy of the speed of light,  it is  evident that observations are always made in 
flat-frames.  
The term $e^{\xi_{f}}$
defines the flat-frames involved in our observations,  
and 
 is defined by   $\omega_{l}$, the expected frequency shifts   for  Keplerian 
 orbital velocities 
 predicted from     photometry:
   \begin{equation}
  e^{\xi_{f}}(r) =   \frac{\omega_{l} (r)}{\omega_{o}}. 
  \label{turq}
     \end{equation}
    Keplerian rotations are our best estimates of what would be observed 
    in the absence of relative
    curvature  effects,  as evidenced by the Solar System orbital velocity profiles which do not require dark matter.

According to the convention established in Eq.~\ref{XO},  the Lorentz mapping term   is defined by the   curved $2$-frame into flat $2$-frame  
\begin{equation}
 (e^{ \xi_2} )^2= \frac{e^{\xi_{f}}}{ e^{\xi_{c}}  }. 
\end{equation} 
The term $e^{\xi_{c}} $  is defined  as per  Eq.~\ref{eq:11}.   

 The  transformation  $v_2$---found to be the most robust across the sample---is the 
a ratio of the hyperbolic Lorentz boost evaluated at the limit of the data, $v_\tau$,  to the same boost at each  radii,  $v(r)$,  
  \begin{equation}
 v_2   = \frac{v_\tau}{v(r)}.  
\label{eq:hyperbolico}
\end{equation}
   The value at  the limit of the data $v_\tau$  is a good approximation  of the final extent of the baryon profile.  The form of the hyperbolic boost   is

  \begin{equation}
\frac{v(r)   }{c}=\tanh \xi_2= \frac{(e^{ \xi_2} )^2-1}{(e^{ \xi_2} )^2+ 1}.  
\label{eq:hyperbolico}
\end{equation}

  \section{Method }
 \label{fitting}
The   LCM  
rotation curve velocity  prediction,  $v_{rot}$ (Eq.\ref{eq:zonteLCM}),
   is fit to    reported rotation curve data $v_{obs}$,  and the $\xi^2$ is minimized to best match the luminous mass profile to the $v_{obs}$ data.   
We use  the MINUIT minimization software as implemented in the ROOT data-analysis package~\citep{ROOT},   as follows:
\begin{enumerate}
 \item A database of modern estimates of the  luminous mass components  (gas, disk, bulge) are   collected from the  cited references. 
 \item   The associated    Newtonian gravitational potential $\Phi$ is calculated for each component and  summed by superposition  to parametrize Eq.~\ref{eq:weakfield};
  \item The   convolution function,  $v_{lcm}$,  is calculated using a choice of the Milky Way luminous mass profile;
  \item The  fit is  performed, and the procedure  iterated to find the optimal luminous  profile    ($v_{obs}$) which best reflects the rotation curve data  ($v_{rot}$);    
  \item the value for the LCM free parameter $\alpha$ is recorded and added to our parameter space to constrain the ansatz in Eq.~\ref{correl}.
\end{enumerate}

By iterating the luminous mass within bounds imposed by photometry, we  reproduce rotation curve data almost exactly (see Fig.~\ref{galaxiesSmallest}).
The fits  and resulting free parameter space reported here represent this   LCM  prescription.

 
\section{Discussion  \&  Analysis} \label{sample}
 
 \subsection{A constraint to the  Milky Way  and  the LCM free parameter}
\label{MW}  
 
\begin{figure}
\includegraphics[scale=0.4]{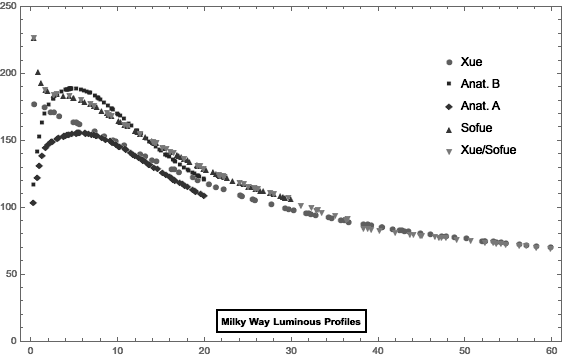} 
\caption{Milky Way Luminous mass   models tested for the LCM mapping construction,  reported in terms of  orbital velocities (km/sec) as a function of radius (kpc).   \label{MWlum}}
\end{figure}

   The LCM is sensitive to both the emitter and receiver galaxies, so the choice of input  Milky Way data is essential.    Since we make observations  from within the system, the  Milky Way luminous mass profile  is  notoriously difficult to determine  ~\citep{Car}.   We have   compared each 
of the  $25$  galaxies in our sample to $(4)$ different, well studied    Milky Way  (MW) luminous mass models; \citet{Sofue}, ~\citet{Xue},  and two models from \citet{Klypin}.  Table~\ref{tab:MWlum}  and  Fig.~\ref{MWlum} compare  the salient features of the   MW  models.   

 The sample  of $25$ galaxies reported here encompasses a large variation in estimication of  the common  physical trends in  the free parameter space.  The complete set of individual  emitter galaxy  results  are 
reported in   \cite{Cisneros2015},  and include
    $\alpha$,  M/L, $r_e$,  the LCM    reduced $\chi^2$,  the  reduced $\chi^2$ values for the DM or  alternative gravity model which were  originally used to  fit  to the same data in the cited texts. The LCM luminous mass results for the 25 galaxies  are   consistent with documented  bounds from photometry \citep{Blok}.

\begin{figure*}
 \centering
\subfloat[][Xue/Sofue synthesis]{\includegraphics[width=0.33\textwidth]{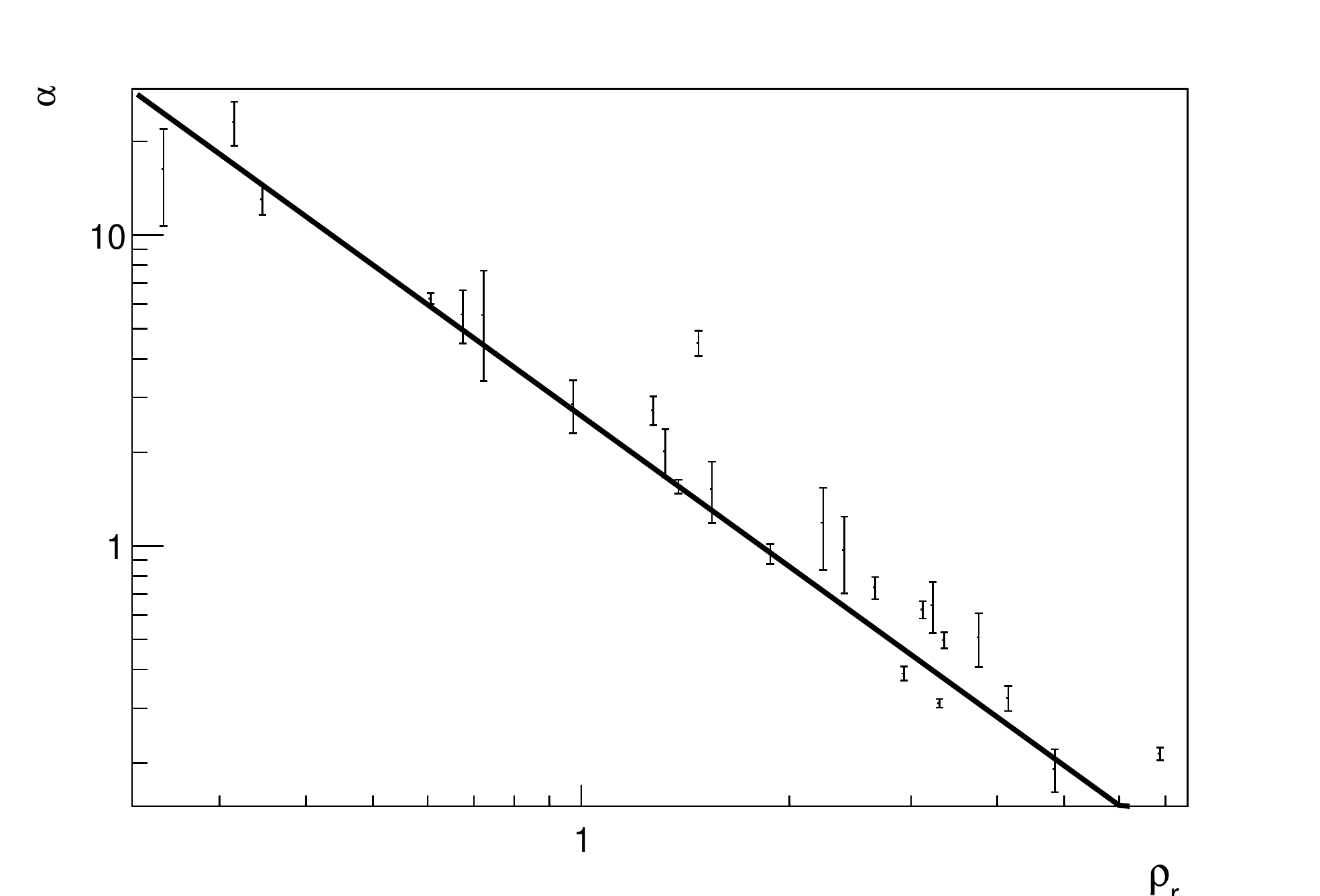}}
\subfloat[][Sofue]{\includegraphics[width=0.33\textwidth]{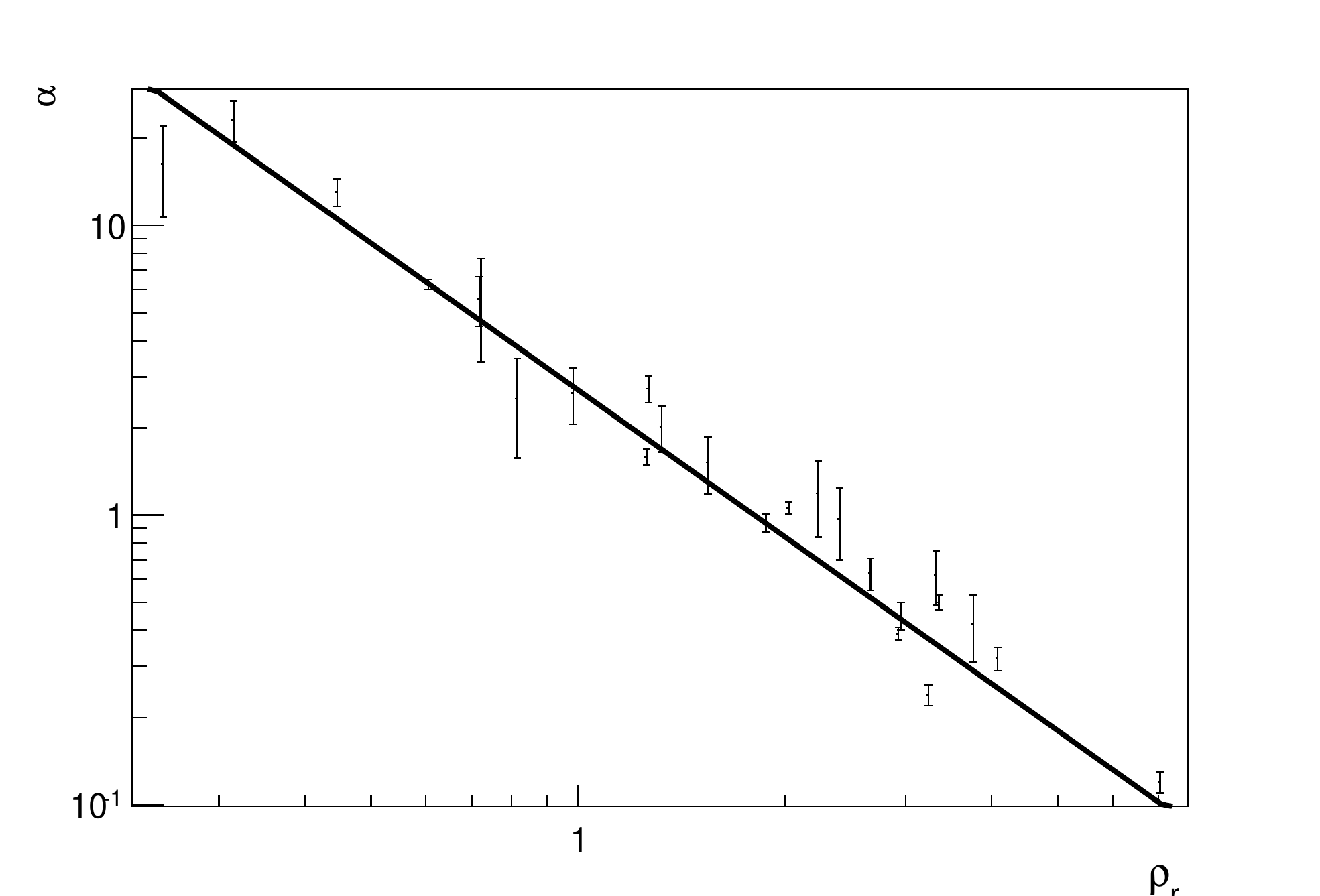}}
\subfloat[][ Xue ]{\includegraphics[width=0.33\textwidth]{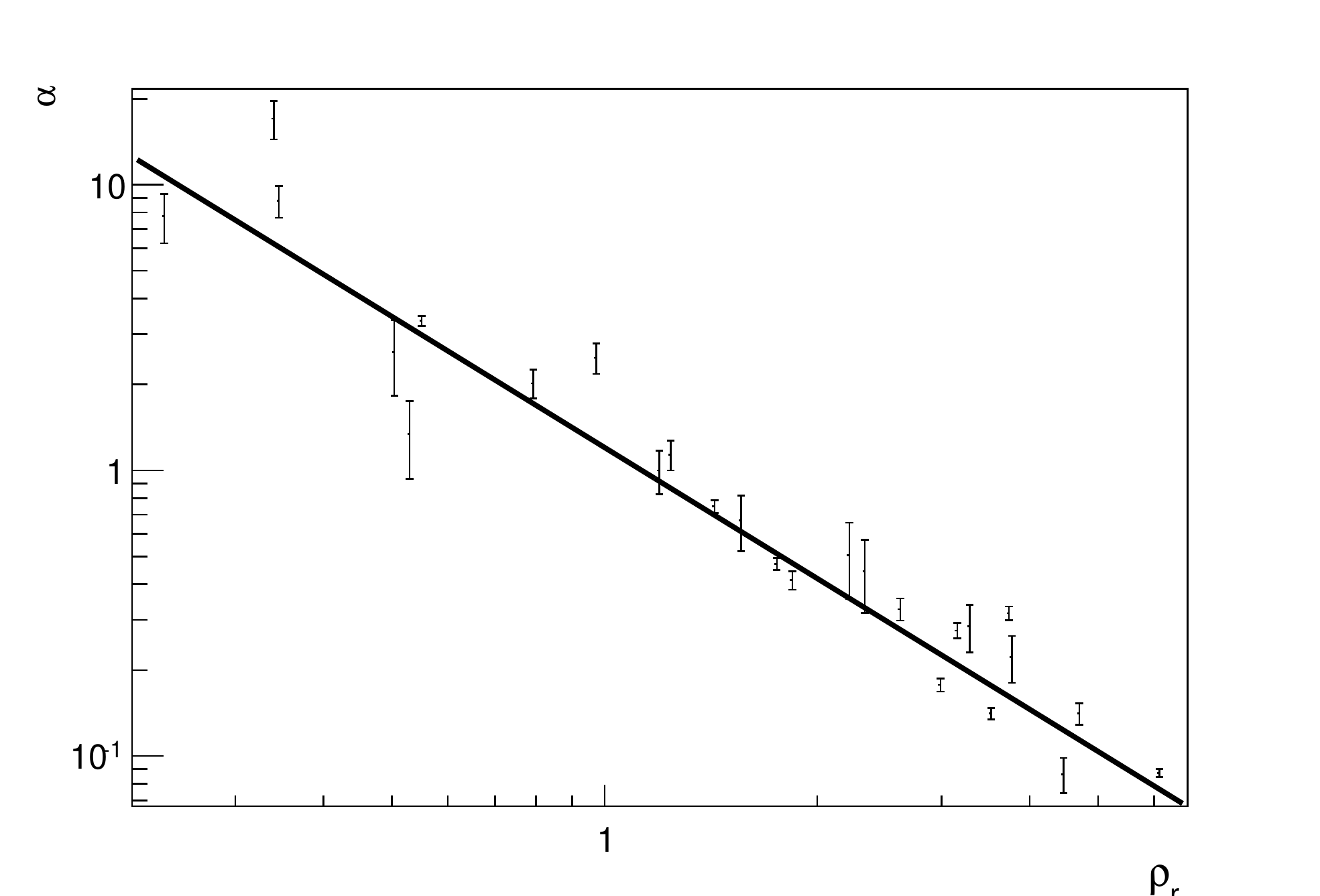}}\\
\subfloat[][ Klypin model A]{\includegraphics[width=0.33\textwidth]{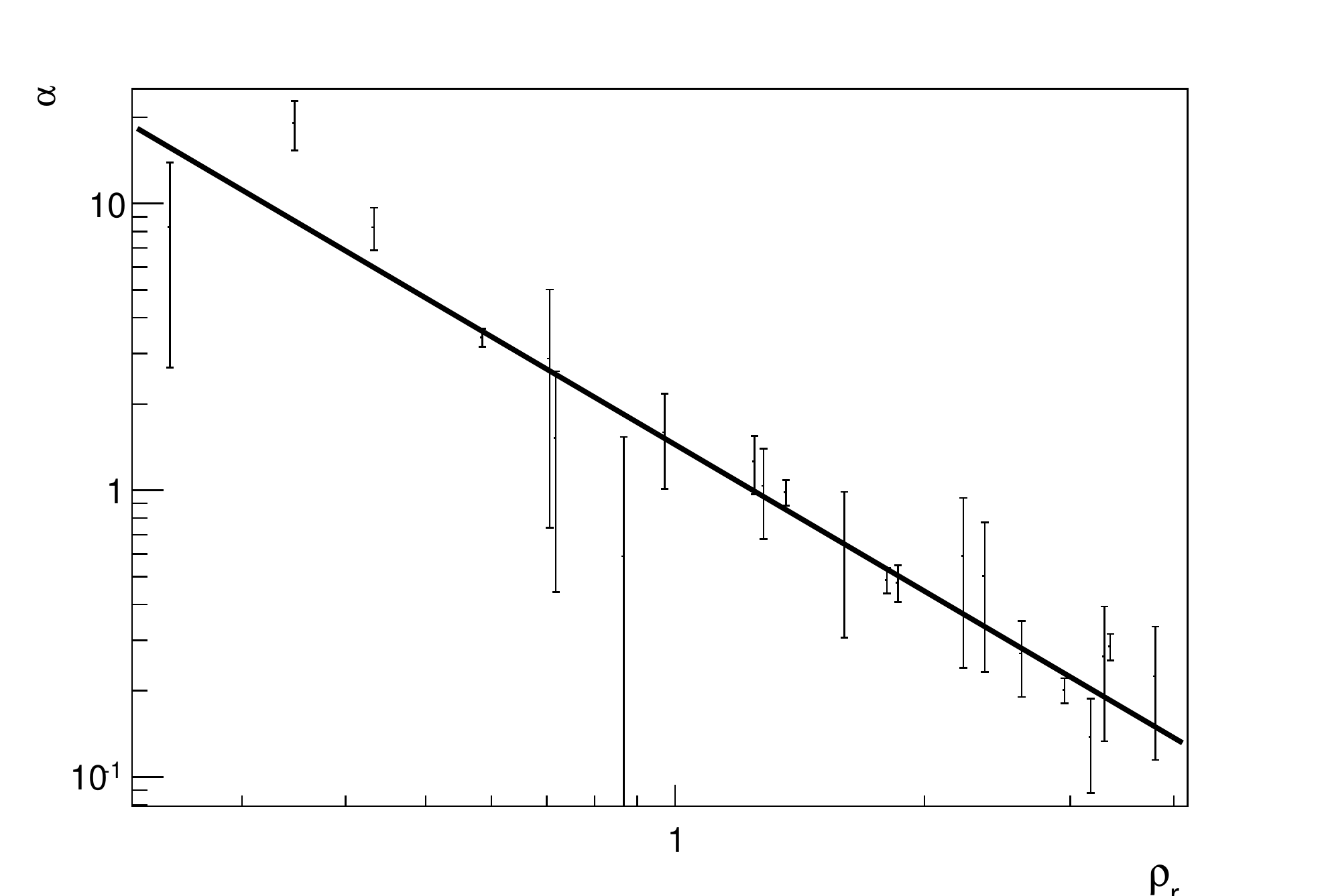}}
\subfloat[][ Klypin model B ]{\includegraphics[width=0.33\textwidth]{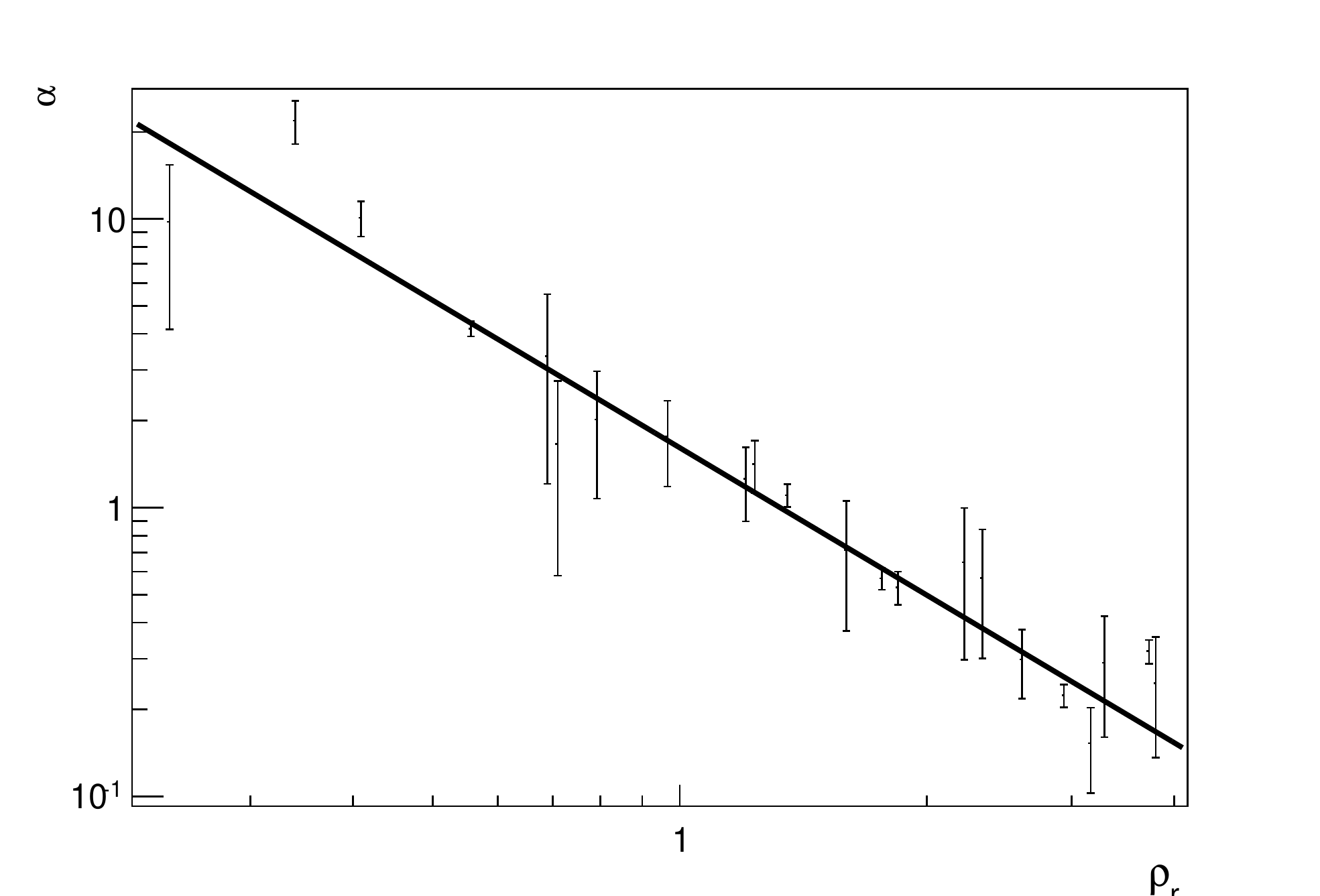}} 
 \\
  \caption{  Log-log plot  of LCM free parameter, $\alpha$,     versus the guess  at   a physical interpretationin in Eq.~\ref{correl} .   
 Each dot represents one of the  galaxies in the sample,   for each Milky Way   model.    Errors  are statistical only. Some Milky Way models did not return LCM fits for each galaxy in the sample.  }    
           \label{fig:alphaDistrib1}
\end{figure*}

  In Fig.~\ref{fig:alphaDistrib1}, we show  the $\alpha$ free parameter space 
       for each MW model.  Each dot in the figures represents  one galaxy $\alpha$ fit result, against thet specified  MW.   To test the correlation of $\alpha$ with the ansatz $\alpha = (\rho_1/\rho_2)^{b}$,  in Eq.~\ref{correl},  we  fit the  luminous profiles which  result  in each  LCM fit with a thin disk, and use the resulting  fit value of the  scale length $r_e$ and an integrated total mass $M$ to parametrize $\rho= M/r_e$.   This
process also gives  the mass to light ratio (M/L) reported previously \citep{Cisneros2015}, using modern reports of the distance and   luminosities $L$ 
  (see Fig.~\ref{fig:alphaDistrib1}). 
Plots of   the $\alpha$ distribution versus the guess $\alpha = (\rho_1/\rho_2)^{b}$ for each MW choice are shown in   in  Fig.~\ref{fig:alphaDistrib1}.   The
    exponent $b$ found for each MW   choice appears to be  constant. Interesting, but beyond the scope of the current work, is that the $b$ exponents for the MW models in our sample are   very similar to the exponents  commonlly reported for dark matter halo density profiles~\citep{NFW}.  Iin a future blind study 
     the LCM free parameter $\alpha$ will be tested as a constant   for a given MW choice against a new sample. Currently, the  most robust MW model by far across the current sample   is  the     \citet{Sofue} MW,  which has an associated value of $b=1.64  \pm  0.03$.

The \citet{Sofue} MW model  is by far the most centrally peaked velocity profile within  the first  kpc.    This is an LCM prediction which can be falsified when the Large Event Horizon \citep{} tightens the  photometric constraints on this region of our galaxy. 

 We find that the MW models in our sample which are less centrally dense can not fit the 
  bulge dominated  galaxies  such as NGC 2841, NGC 7814, NGC 7331 and NGC 5055  without artificial bounding on the model's free parameter.  
 Even when   bounding is imposed artificially  in the fitting process, the remain very centrally dense  galaxies such as NGC 7814 which can only be fit  using  the  Sofue MW.   Fits using the  less centrally dense MW models fail completely for this galaxy.     In fact, it is the sensitivity  to MW choice for these type of galaxy which allows degeneracy breaking of  our central MW baryon gradient. 
  In this paper,  all rotation curve figures   
  reflect the hybrid MW choice of the Sofue stellar bulge, and the \citet{Xue} stellar disk.   The Xue stellar disk is used due to it's high radial extent which allows LCM  galaxy fits  out to $60$ kpc in reported rotation curve data. 
  The stellar disks of all four MW model reported here are asymptotically very similar. 
 
\begin{figure*}
  \thispagestyle{empty}
 \centering
\subfloat[][M 31,  Ref.~7]{\includegraphics[width=0.23\textwidth]{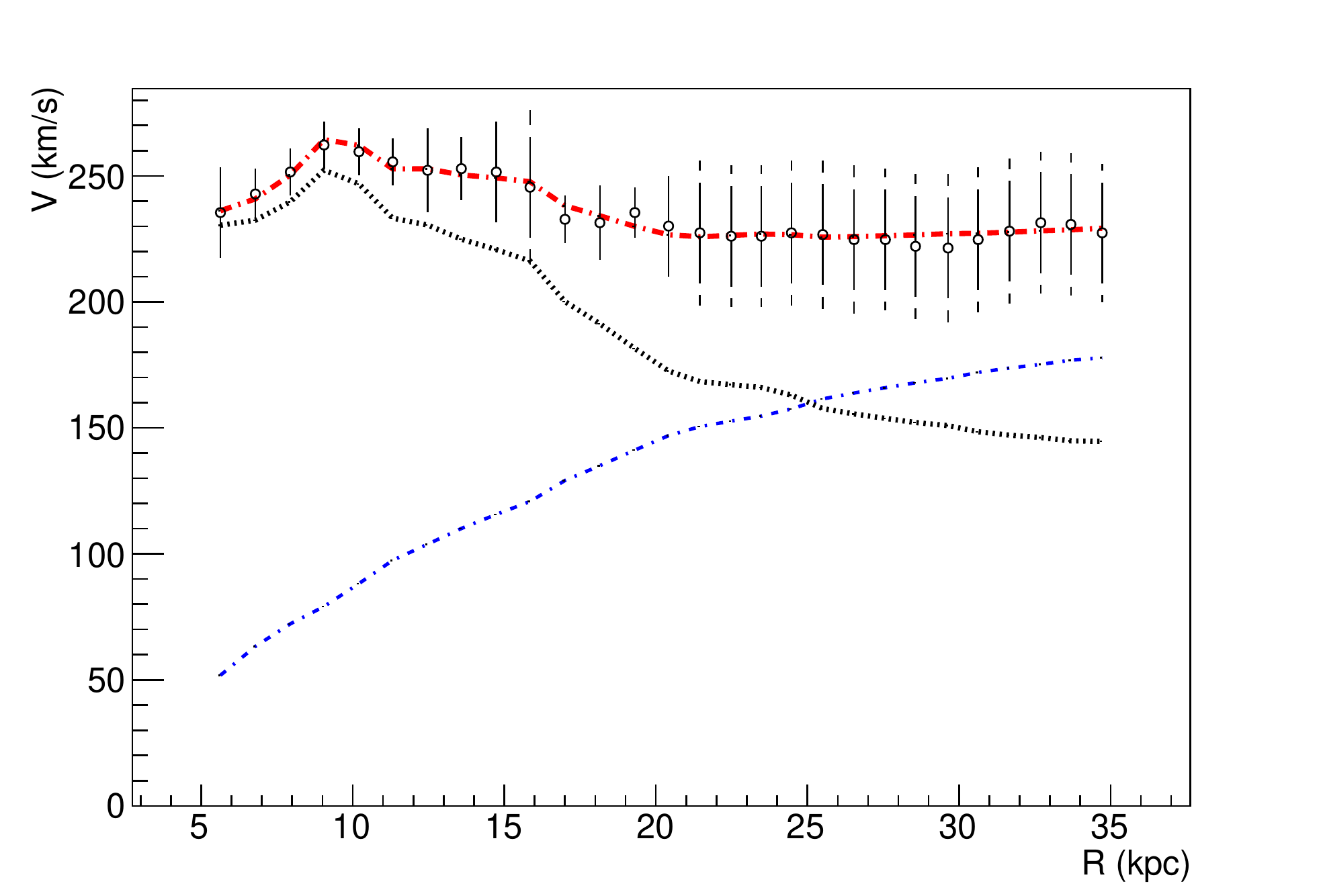}}
\subfloat[][NGC 5533, Ref.~5]{\includegraphics[width=0.23\textwidth]{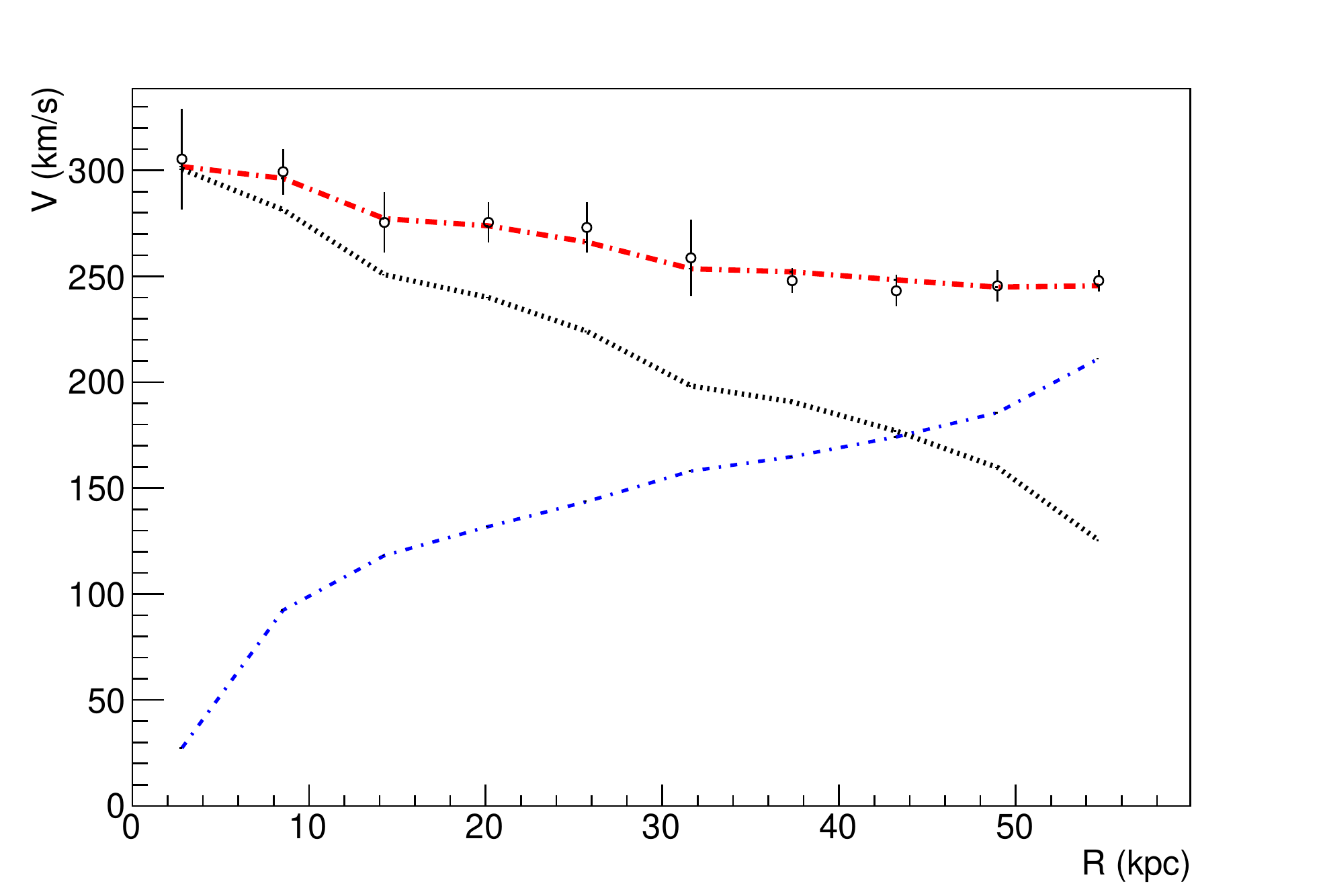}}
\subfloat[][ NGC 7814,  Ref.~6 ]{\includegraphics[width=0.23\textwidth]{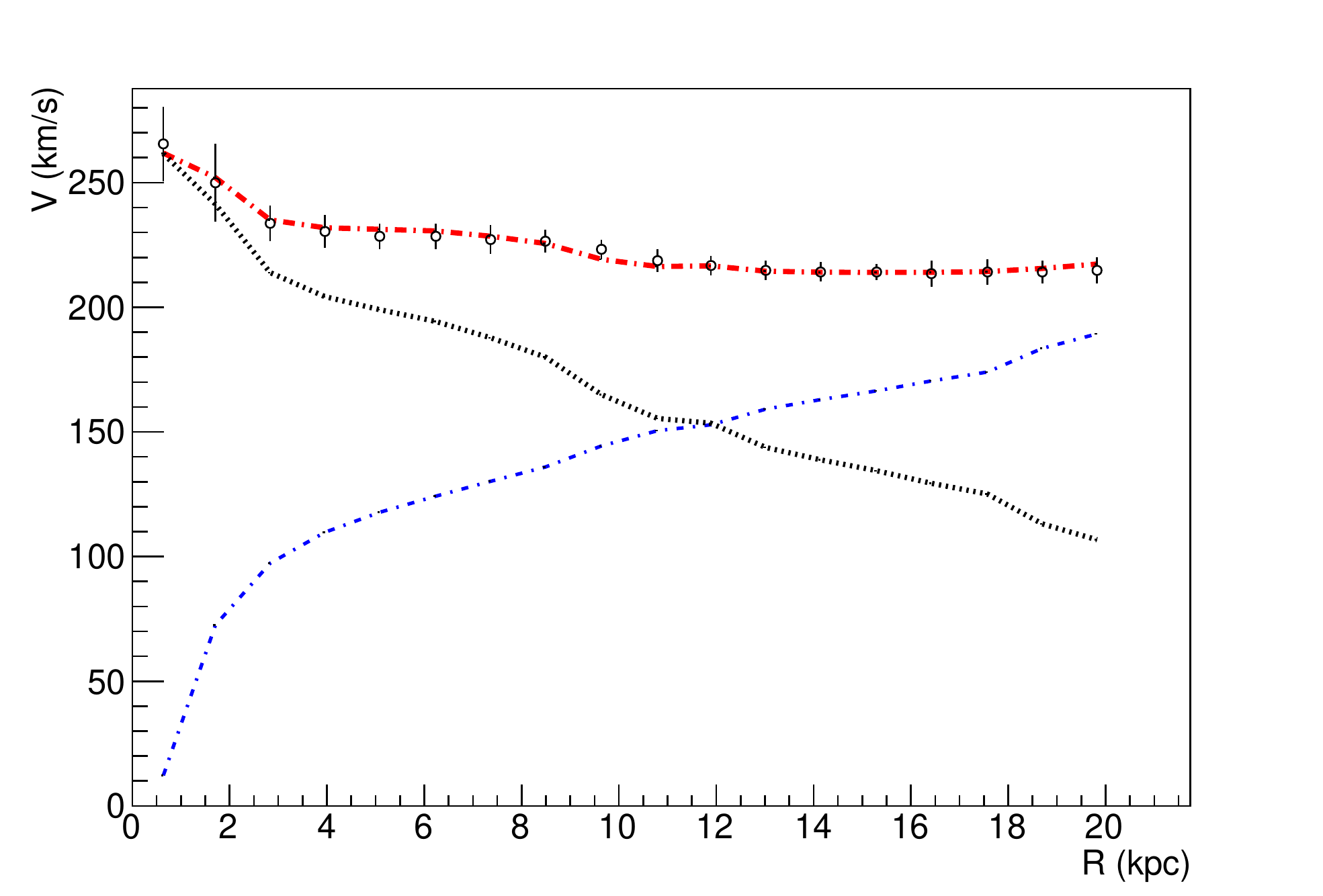}}
\subfloat[][ NGC 891, Ref.~6  ]{\includegraphics[width=0.23\textwidth]{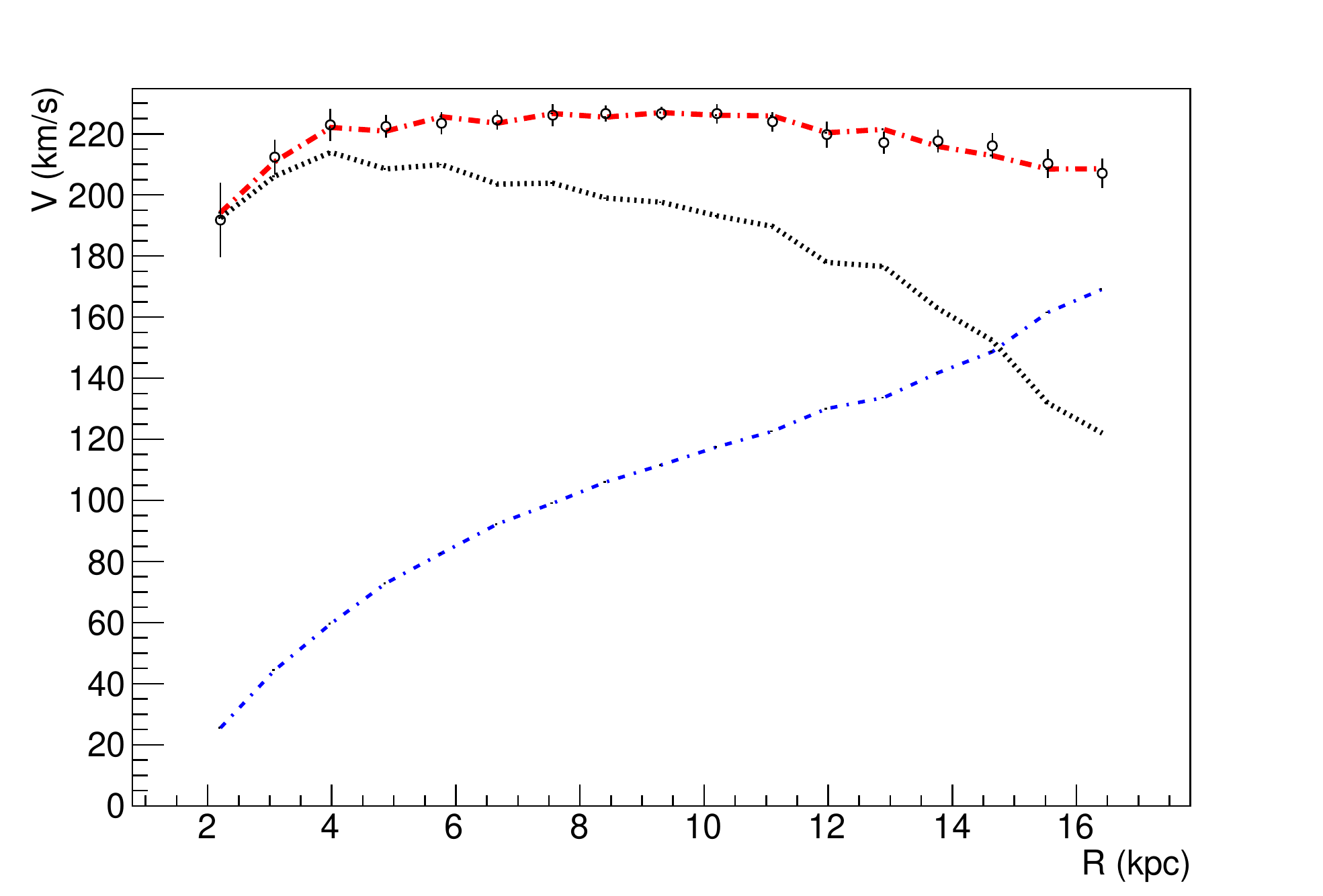}}
\\
\vspace{0.1cm} 
\subfloat[][ NGC 2841, Ref.~4 ]{\includegraphics[width=0.23\textwidth]{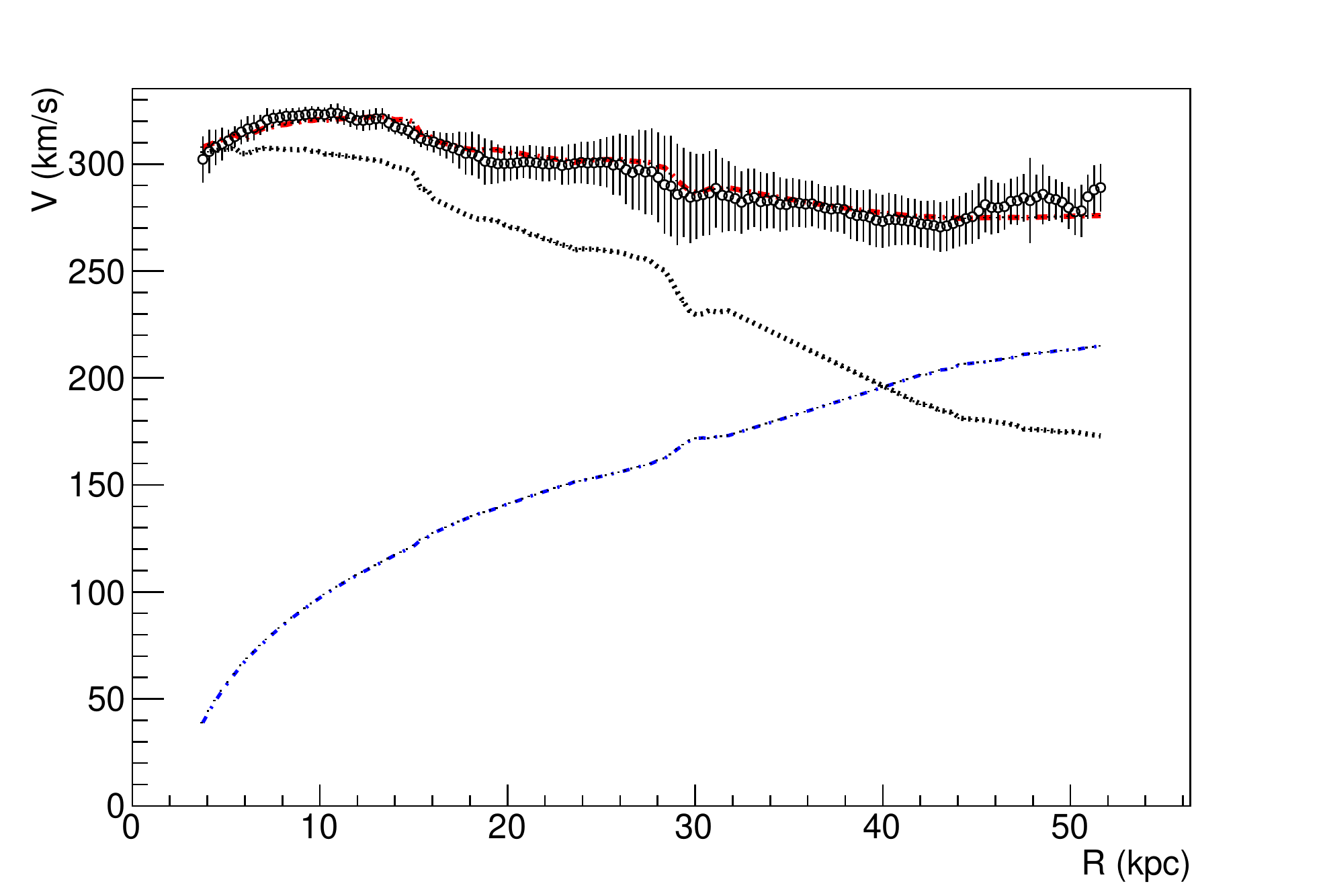}} 
\subfloat[][NGC 7331,~Ref.~4 ]{\includegraphics[width=0.23\textwidth]{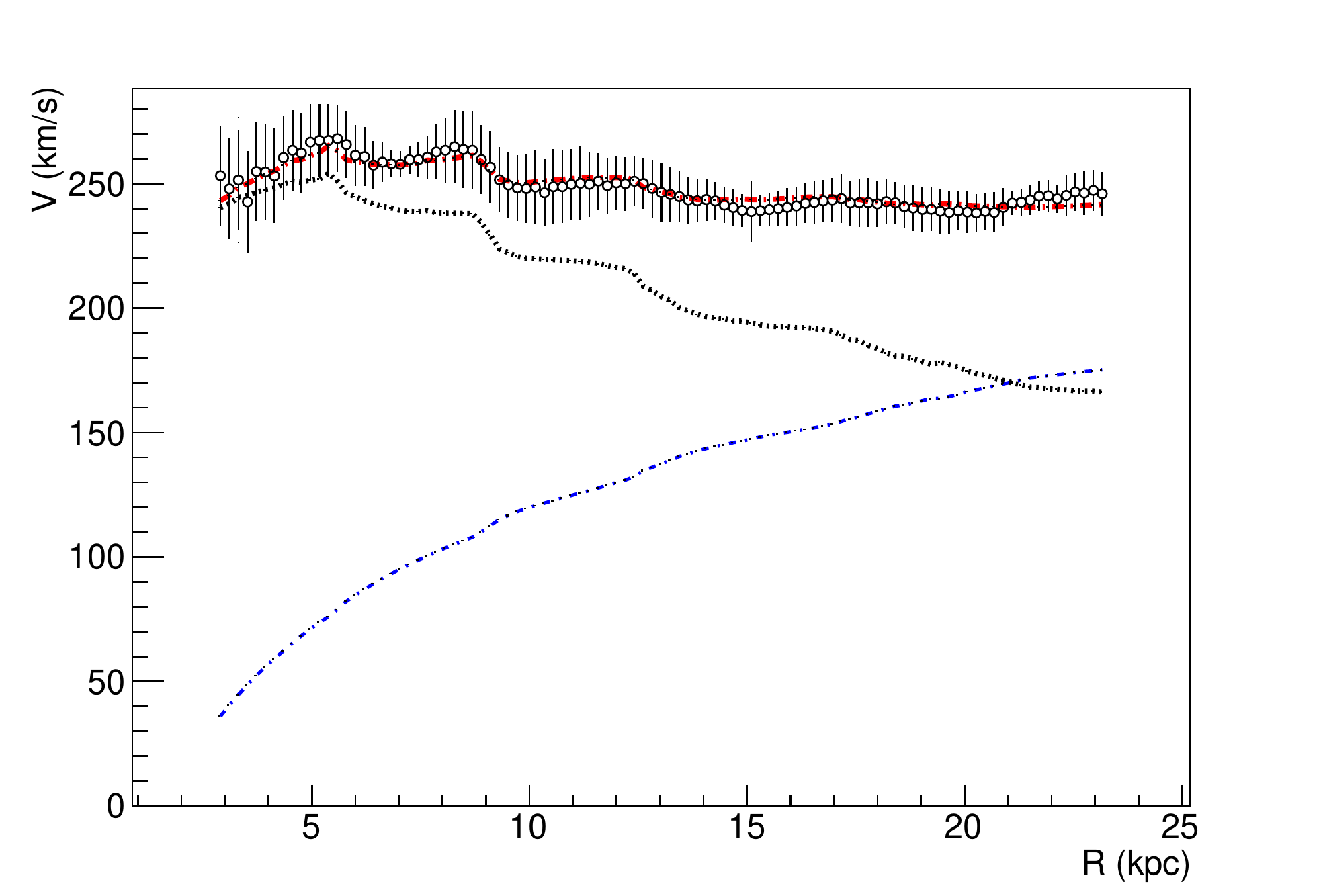}} 
\subfloat[][NGC 3521, Ref.~4  ]{\includegraphics[width=0.23\textwidth]{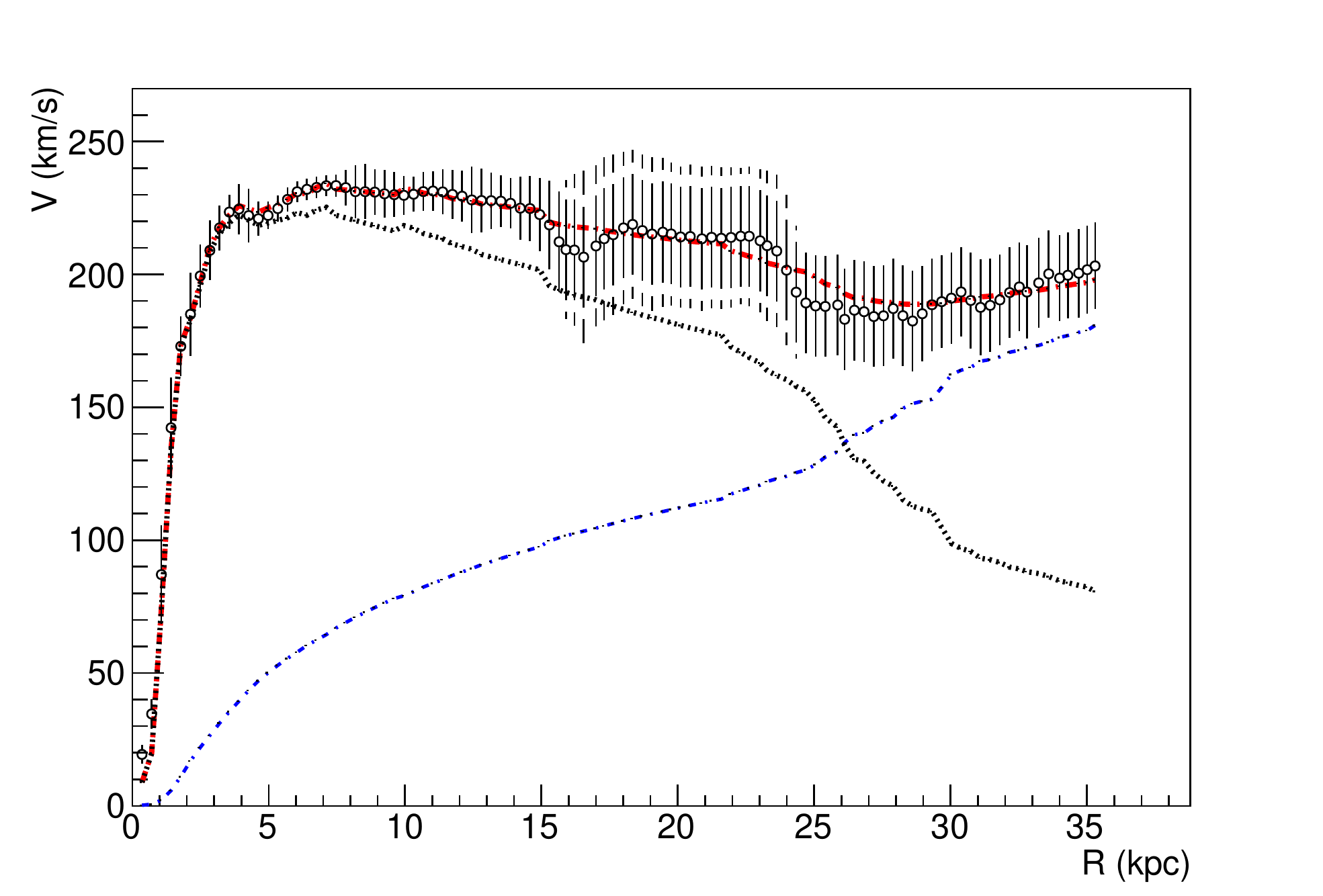}}
\subfloat[][ NGC 5055, Ref.~2   ]{\includegraphics[width=0.233\textwidth]{n5055LCM}} 
\\
\vspace{0.1cm} 
\subfloat[][ NGC 4138, Ref.~5]{\includegraphics[width=0.233\textwidth]{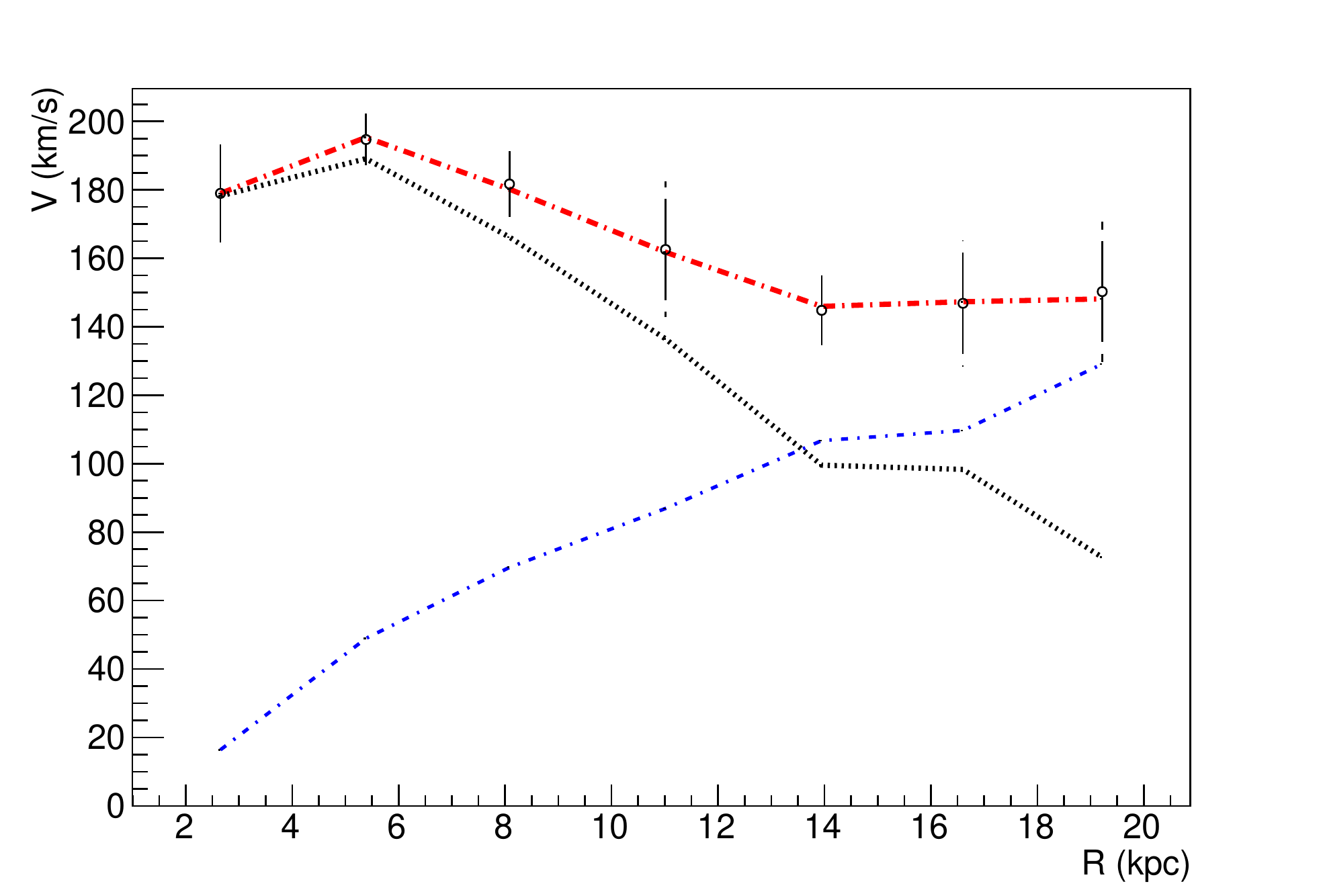}} 
\subfloat[][ NGC 5907,  Ref.~5 ]{\includegraphics[width=0.233\textwidth]{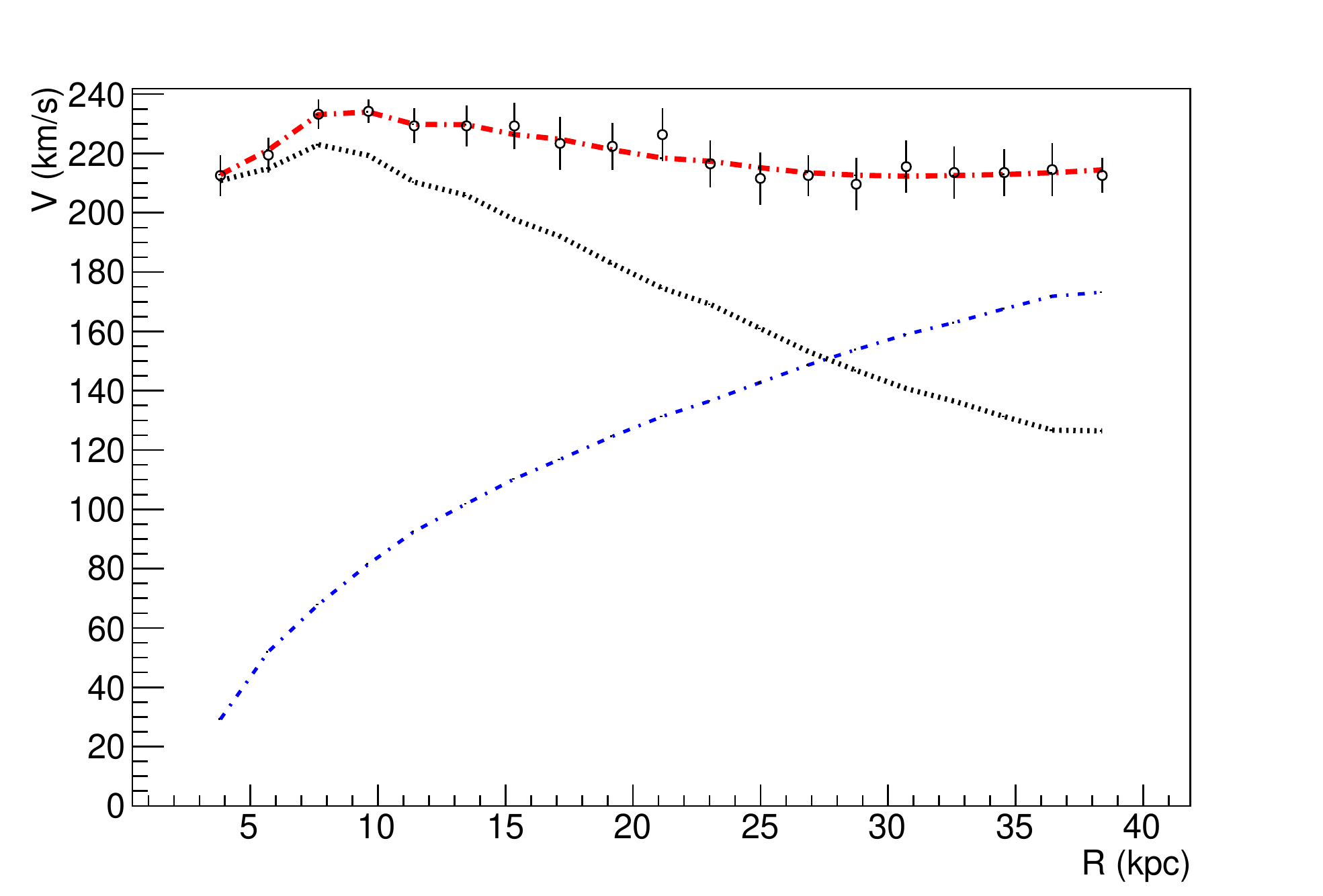}}  
\subfloat[][NGC 3992, Ref.~5]{\includegraphics[width=0.233\textwidth]{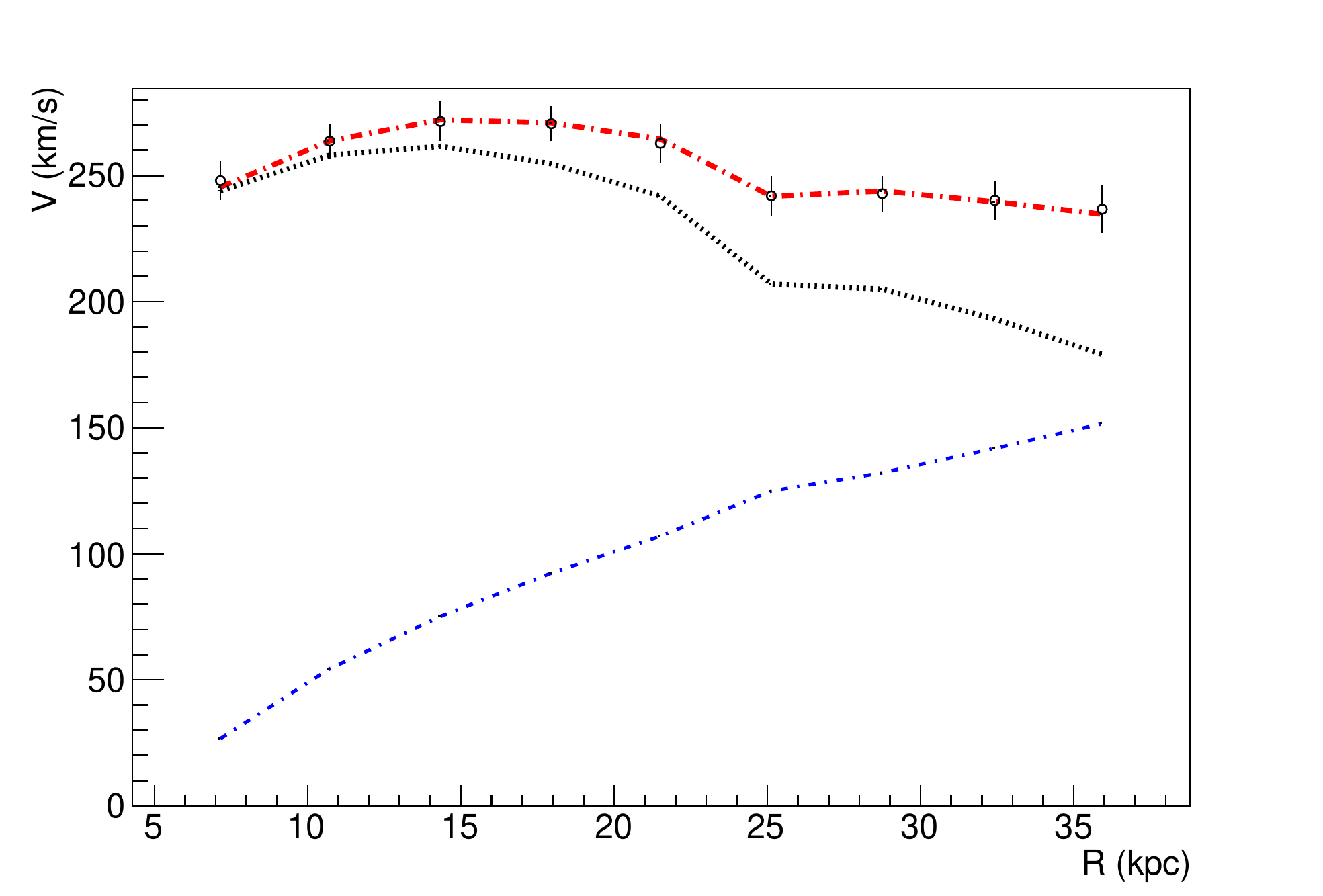}}
\subfloat[][NGC 2903, Ref.~5 ]{\includegraphics[width=0.233\textwidth]{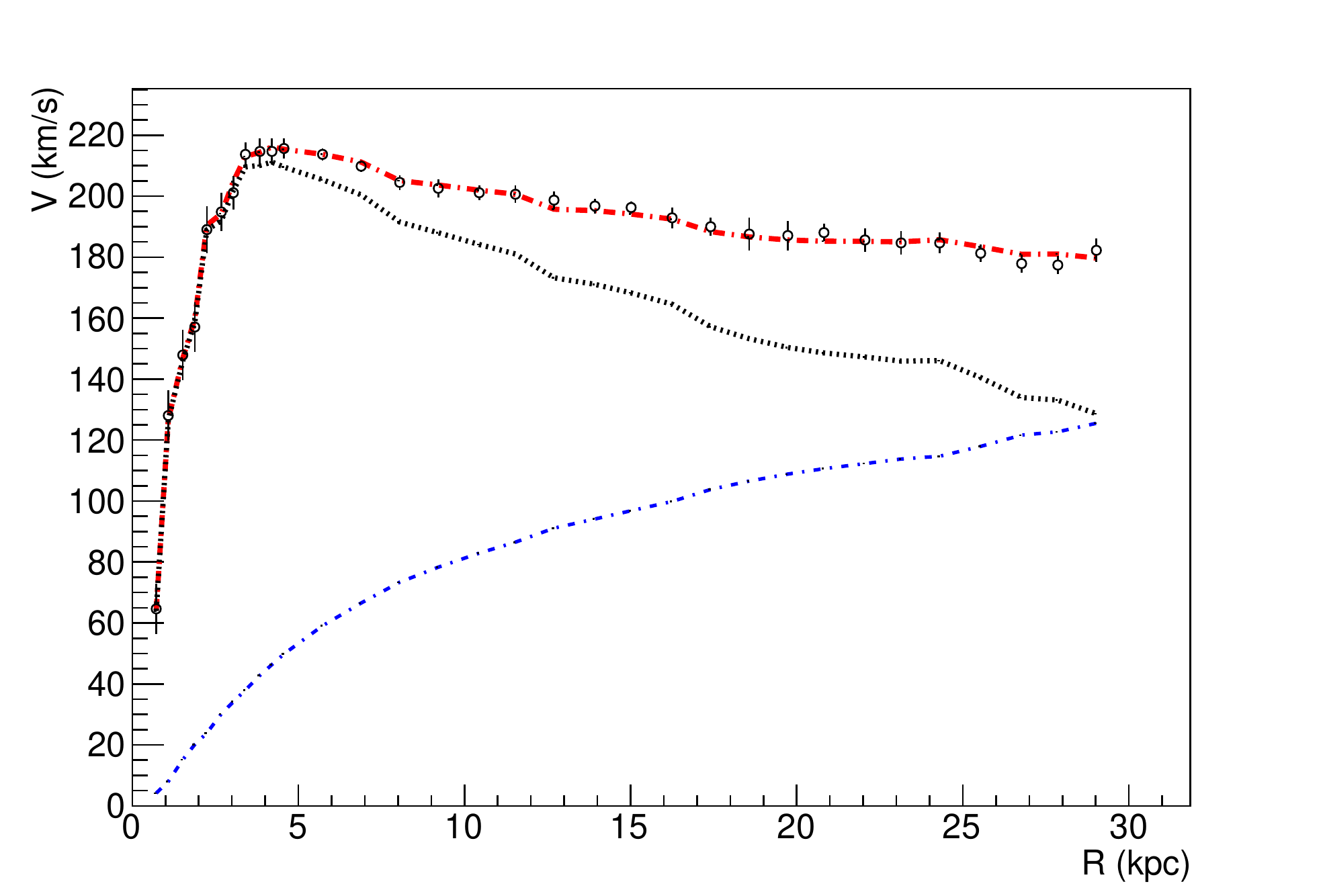}}
 \\
\vspace{0.1cm}  
\subfloat[][NGC 6946, Ref.~5]{\includegraphics[width=0.233\textwidth]{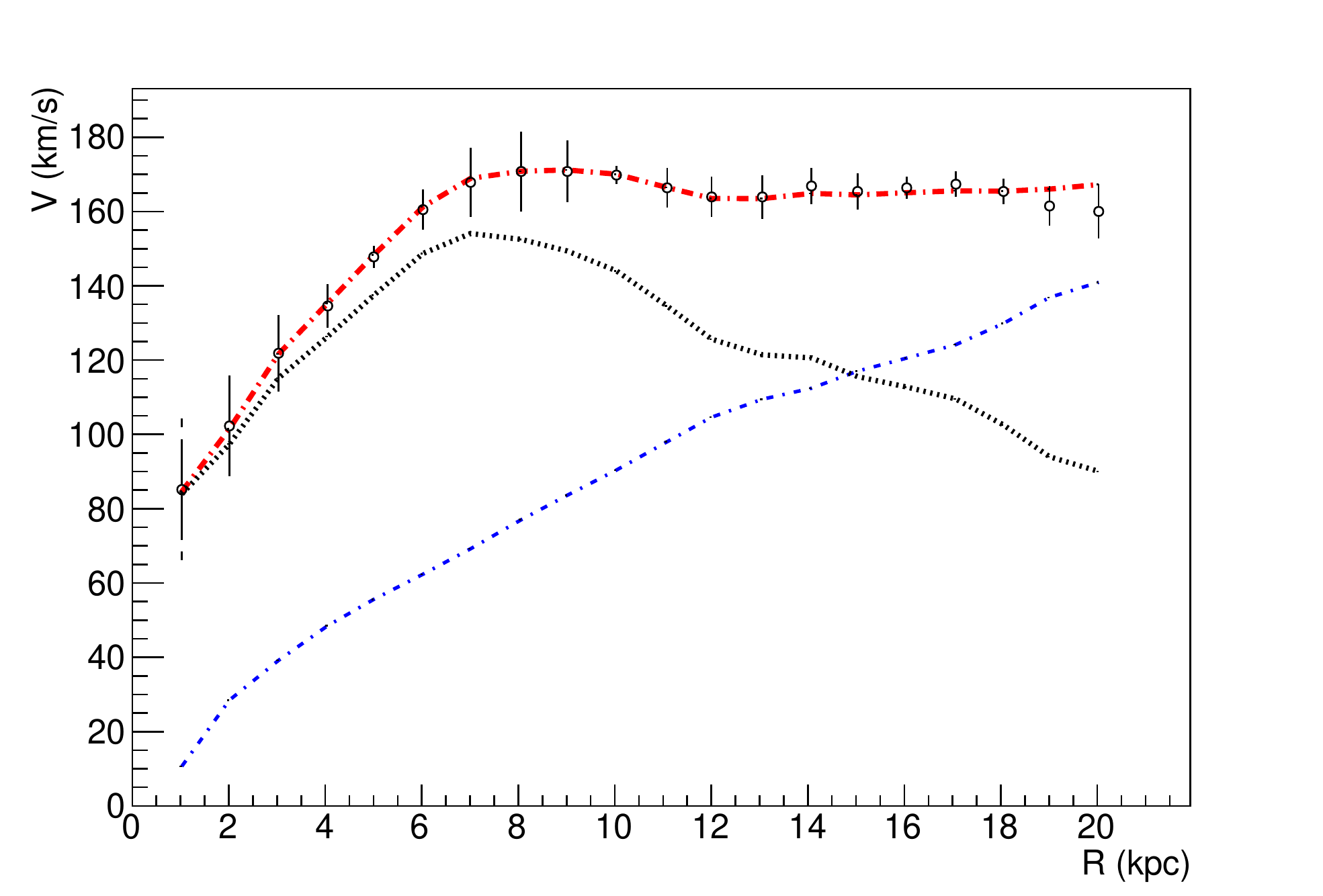}}
\subfloat[][ NGC 3953, Ref.~5 ]{\includegraphics[width=0.233\textwidth]{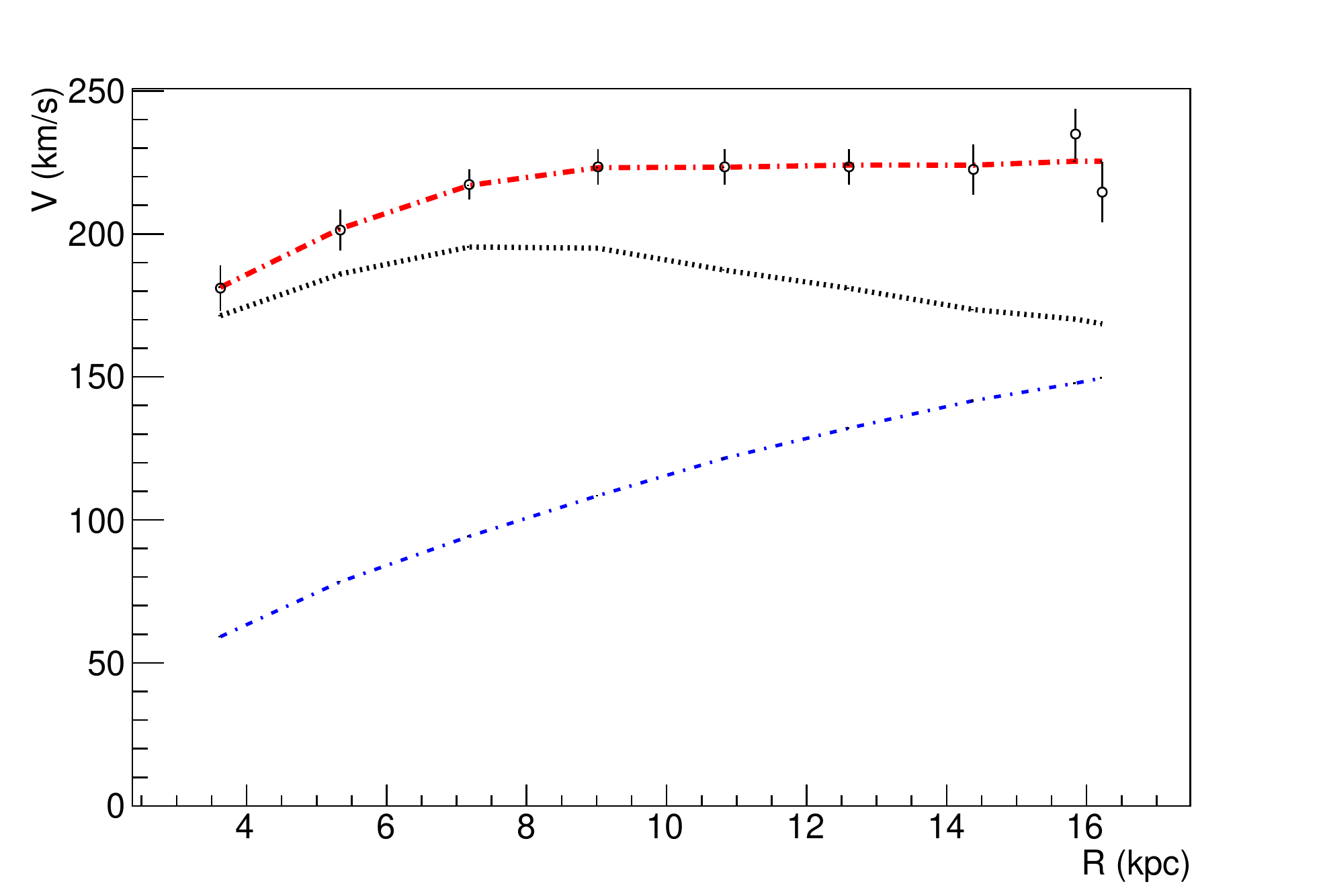}}
\subfloat[][ UGC 6973, Ref.~5 ]{\includegraphics[width=0.233\textwidth]{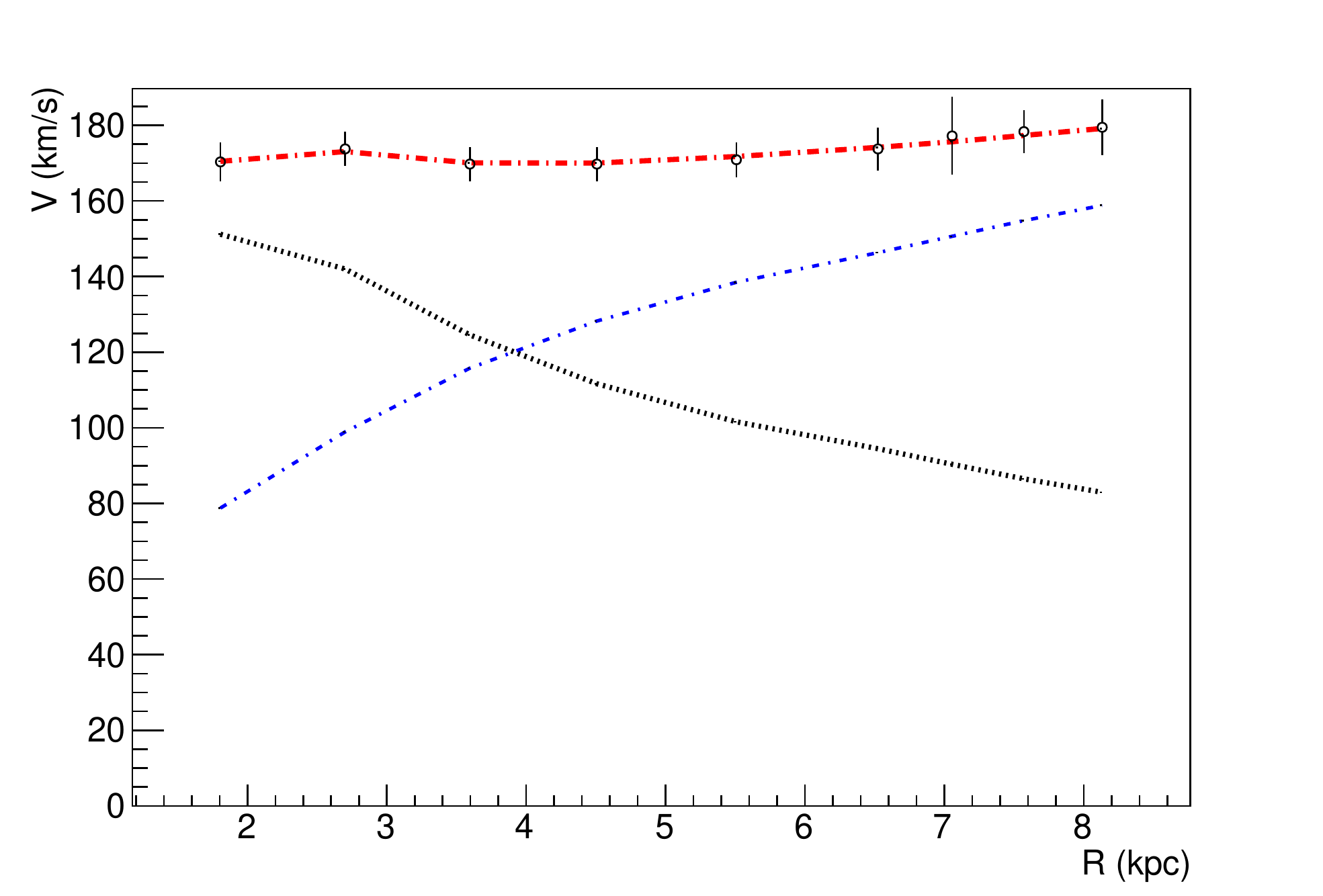}}
\subfloat[][ NGC 4088, Ref.~5]{\includegraphics[width=0.233\textwidth]{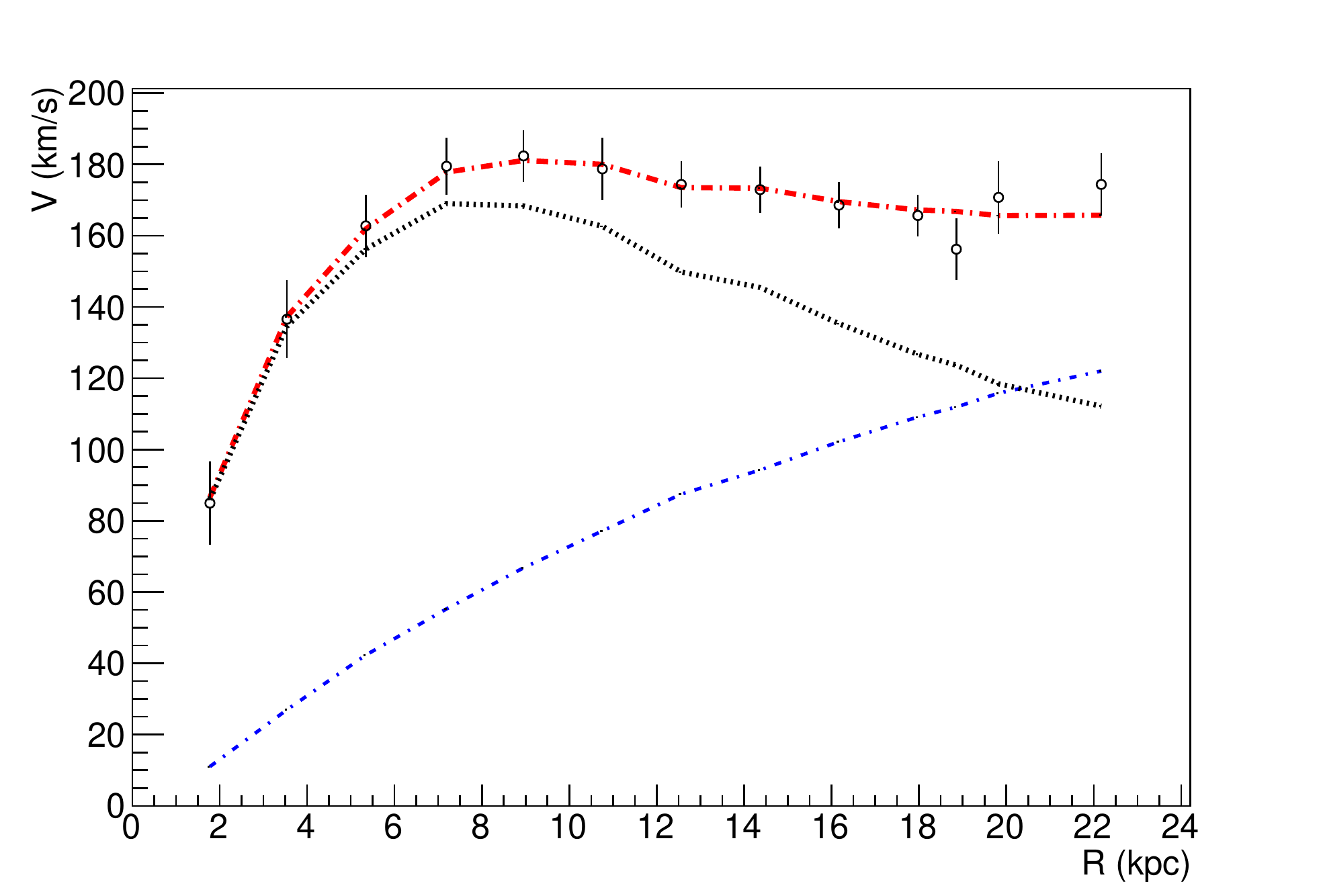}} 
 \\
\vspace{0.1cm} 
\subfloat[][ NGC 3726, Ref.~5]{\includegraphics[width=0.233\textwidth]{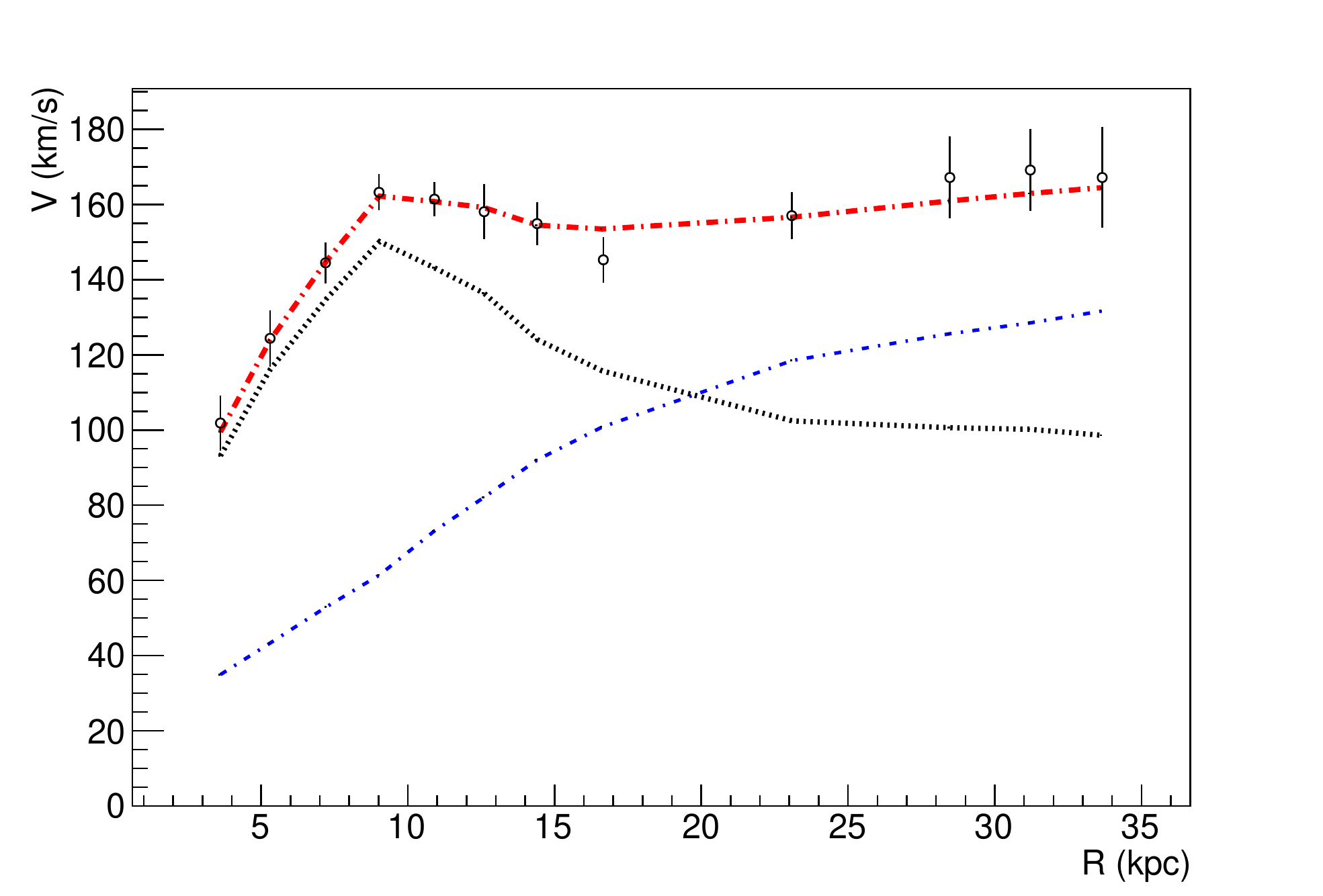}} 
\subfloat[][ NGC 2403, Ref.~2]{\includegraphics[width=0.233\textwidth]{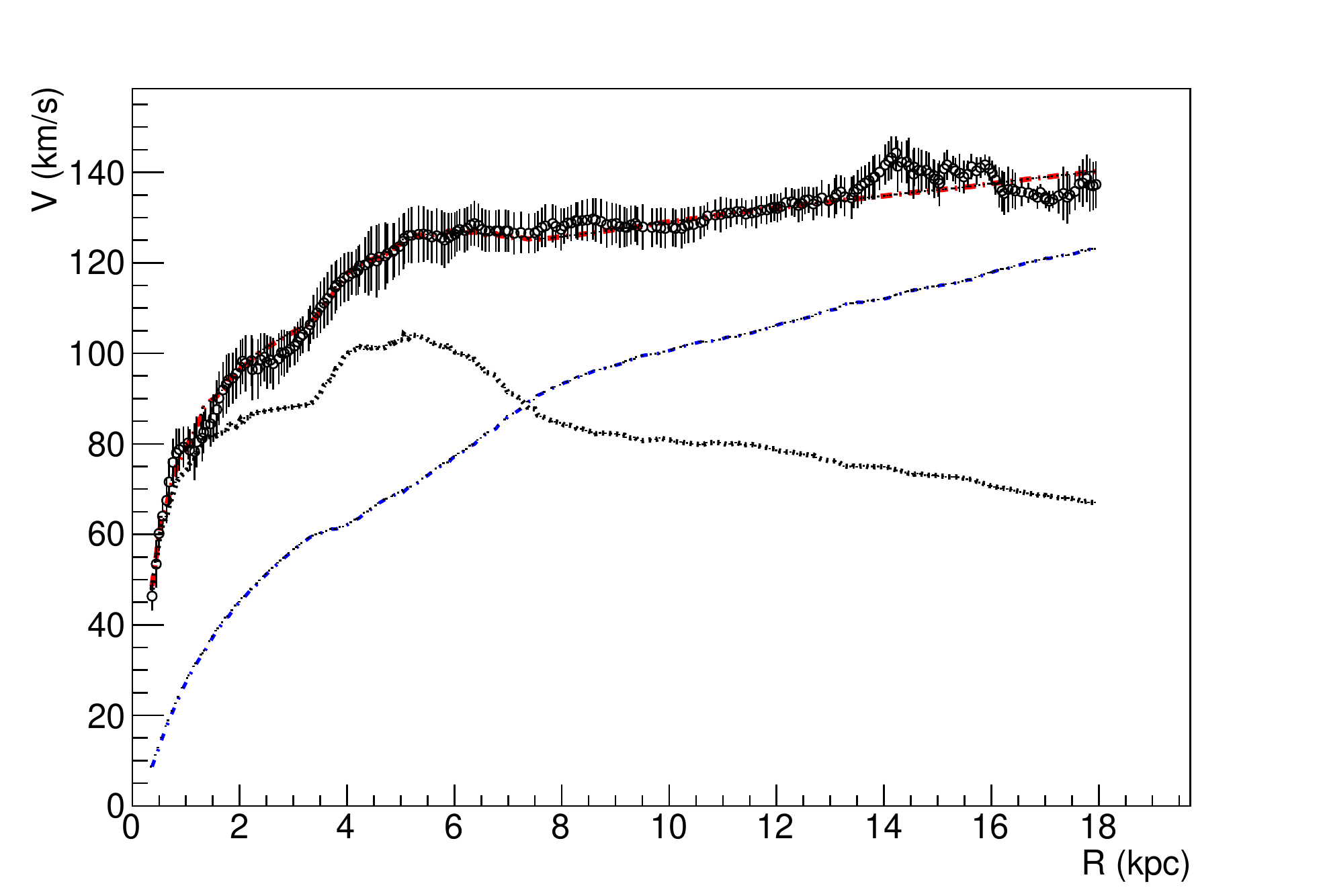}}
\subfloat[][ NGC 3198, Ref.~2]{\includegraphics[width=0.233\textwidth]{n3198cLCM}}  
\subfloat[][M 33, Ref.~3]{\includegraphics[width=0.233\textwidth]{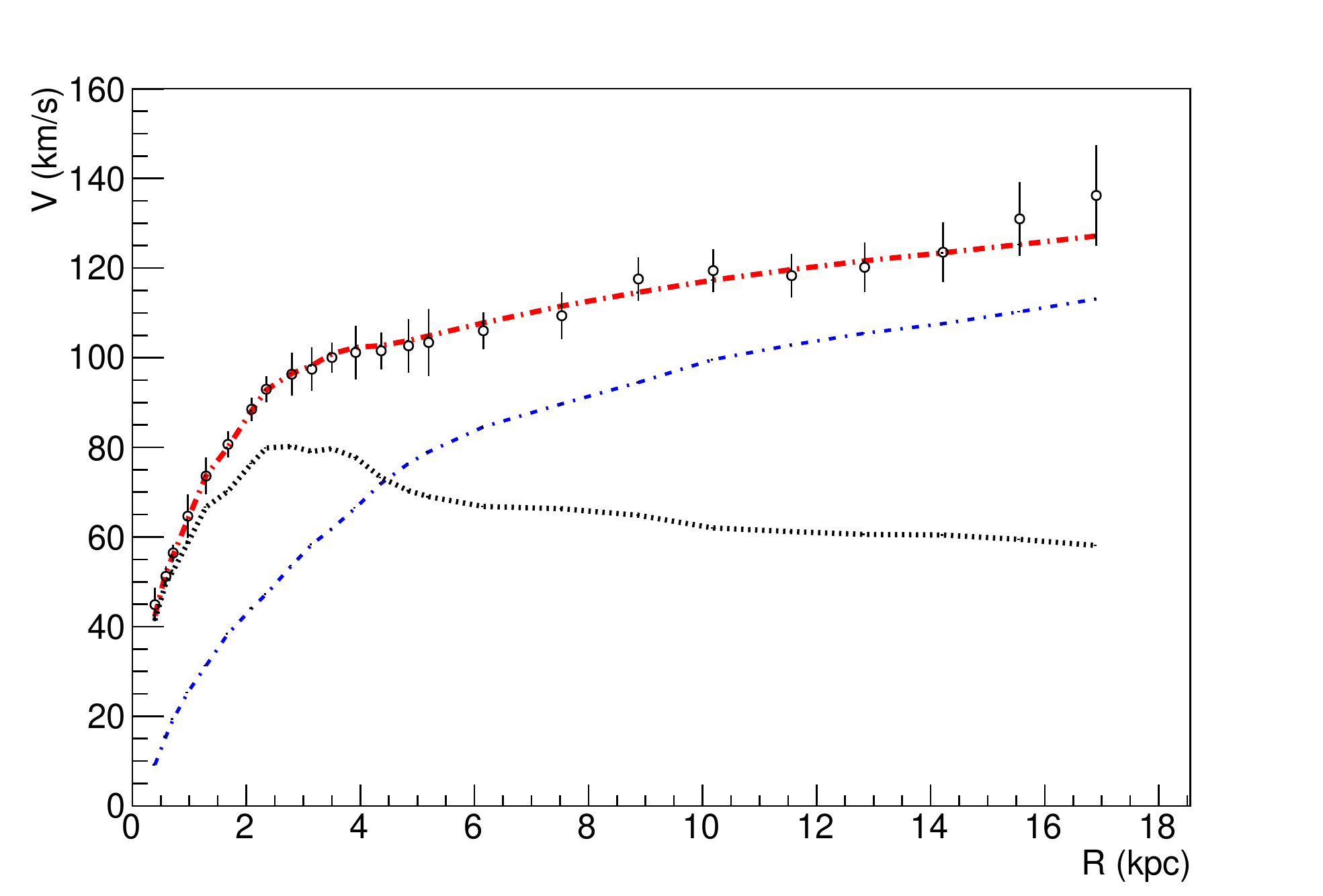}}  
 \\
\vspace{0.1cm}   
\subfloat[][ F 563-1, Ref.~1]{\includegraphics[width=0.233\textwidth]{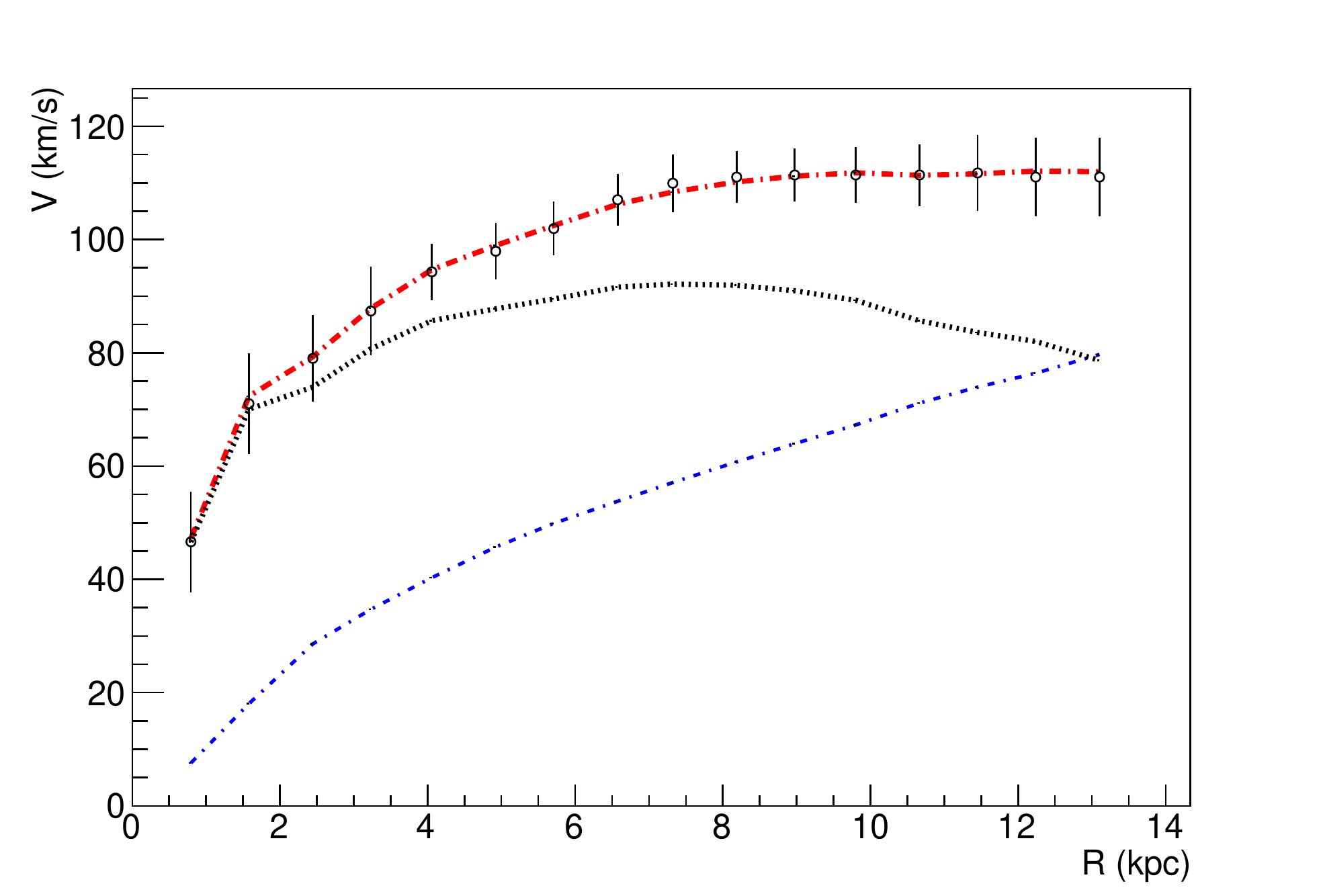}} 
\subfloat[][ NGC 7793, Ref.~8]{\includegraphics[width=0.233\textwidth]{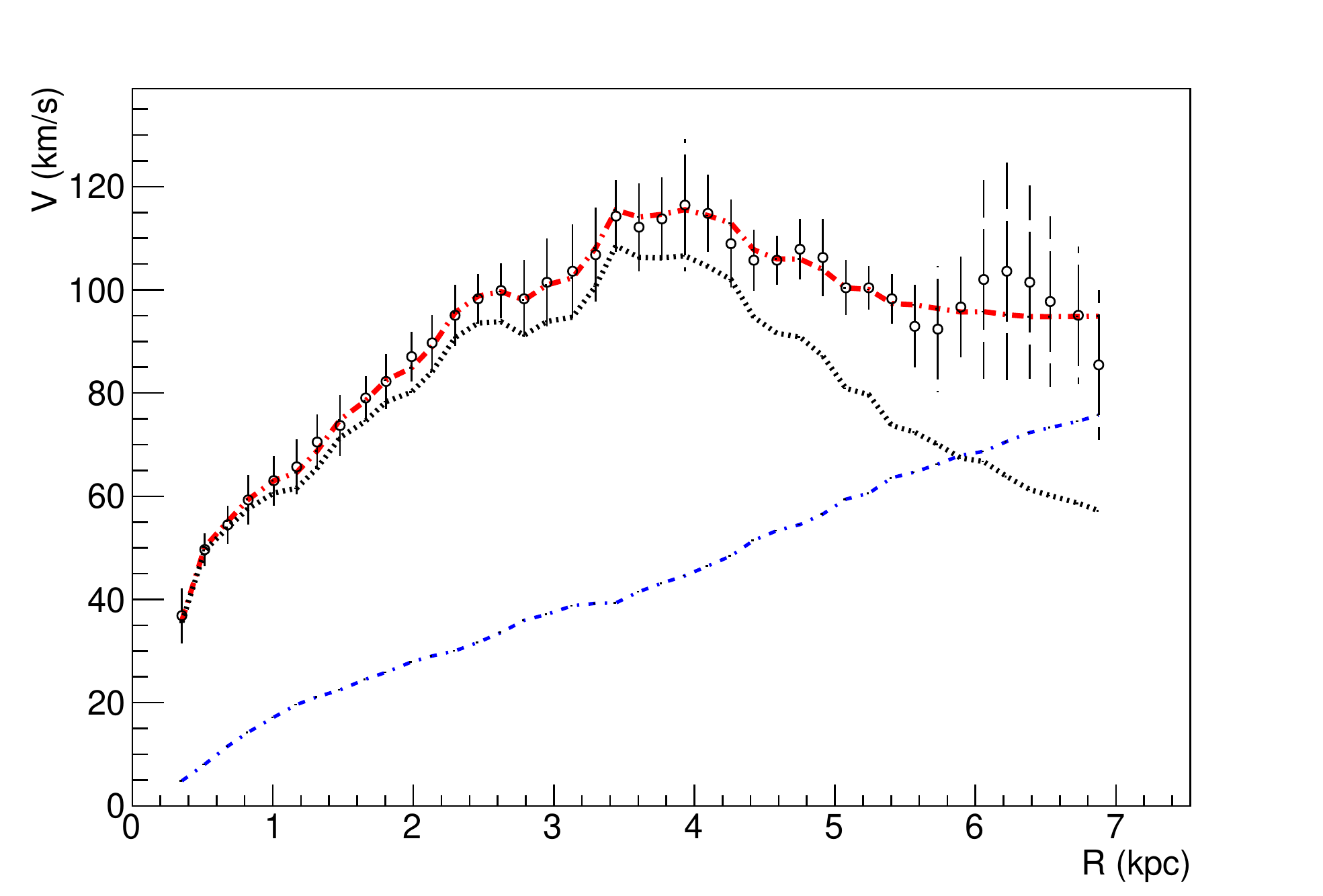}} 
\subfloat[][ NGC 925, Ref.~2  ]{\includegraphics[width=0.233\textwidth]{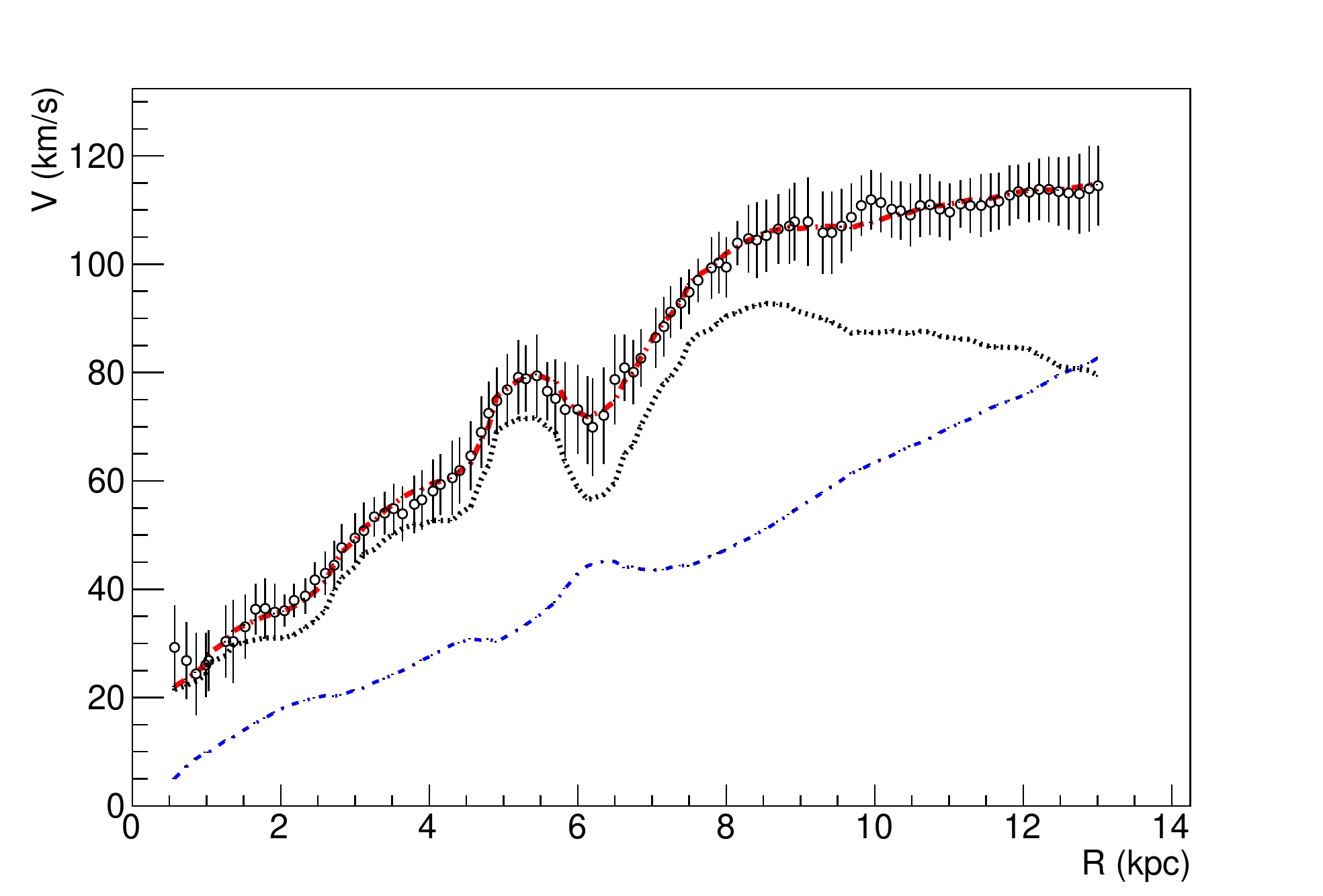}} 
\subfloat[][UGC 128, Ref.~4]{\includegraphics[width=0.233\textwidth]{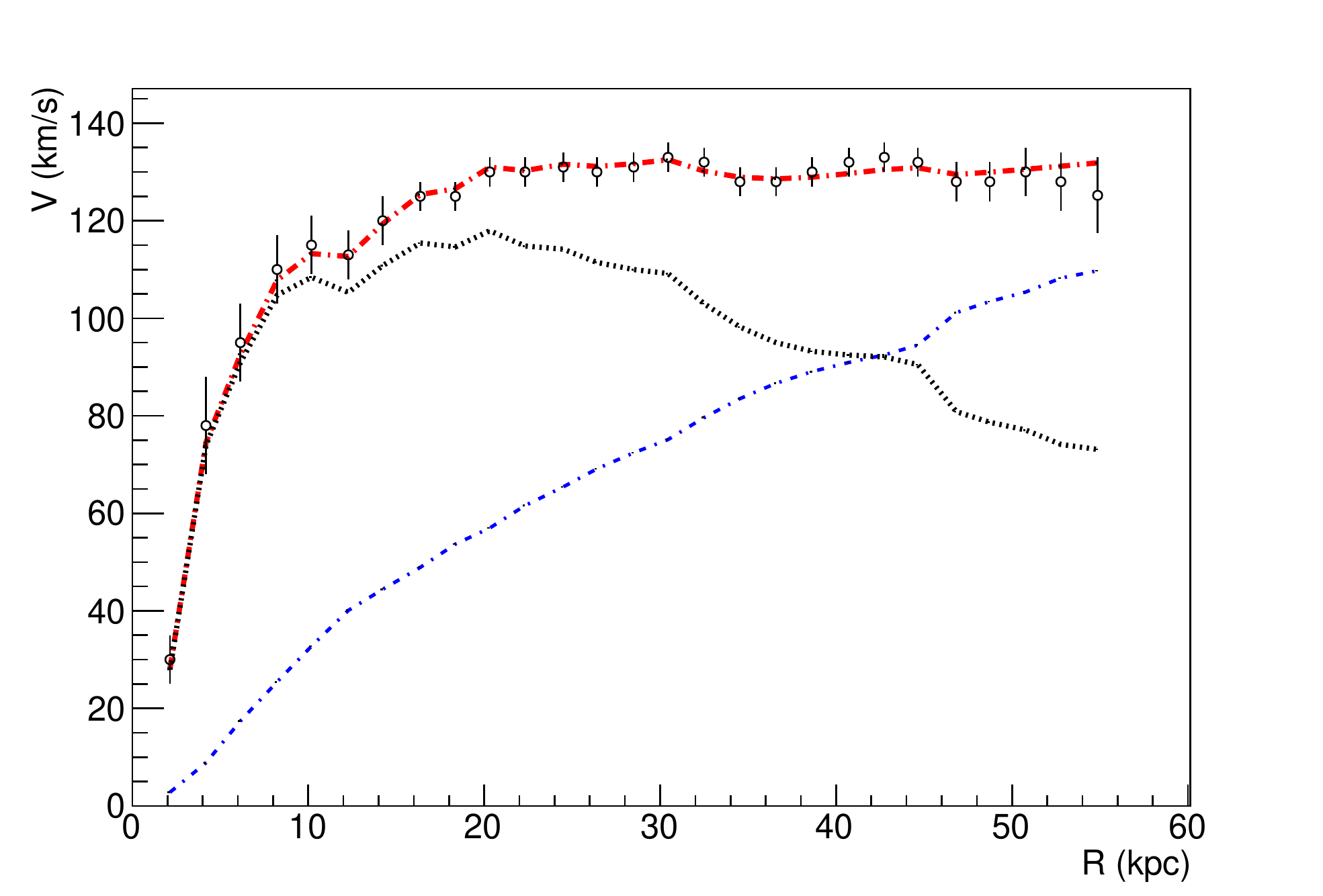}}
  \\
\vspace{0.1cm} 
 \caption{   Circles  with associated error bars are reported rotation curve data,     red dotted-dash lines are LCM fits,   blue dotted-dash line  are   relative curvature,   and     black dotted line are  luminous mass results.   All fits shown are in comparison to the \cite{Sofue,Xue} Milky Way model.  References:   1. ~\cite{JNav},  2.~\cite{Blok} , 3.~\cite{Cor03}, 4.~\citet{James},   5.~\cite{SanMcGa},
  6.~\cite{Frat},   7.~\cite{Car},  8.~\cite{Dicaire}.    }  
             \label{galaxiesSmallest}   
\end{figure*}

 It is also interesting to note that the two galaxies NGC 891 and NGC 7814 from   \cite{Frat} in  can be compared as an example of galaxies with   very similar enclosed mass at the limit of the data, but  very different distributions of that mass. In Fig.~\ref{galaxiesSmallest}, it is clear that the slope of the relative curvature contributions for these two galaxy reflect 
 the  different density gradients.    The ulta-sensitivity of 
 NGC 7814 to small changes in the input luminous mass profile (within a tolerance of    $\pm10$s of  m/sec ) is not mimicked by the NGC 891 fits,  which is  robust to  a  range of      MW choices.   We believe  the sensitivity of NGC 7814  fits, and for similar  galaxies,   is due to   when a galaxy being observed is very similar to the    curvature of our own MW in some region.   For NGC 7814, that region is the central one kpc.

      \begin{table*}
   \centering
  \begin{minipage}{100mm}
  \caption{ Milky Way Luminous Mass Model results}
\label{tab:MWlum}
  \begin{tabular}{@{}llccclrr@{}}
  \hline
 Galaxy     	 & $R_{last} $ &$M_{bulge} $&$M_{disk} $&$ b$ & $r_e$ & $\chi^2_R$ & $ dof $\\ 
 \hline
 Xue/Sofue 	&60 			&$1.8$		&$5.30$ 		&$1.61$    	 &3.74	&13.19& 24\\
 Sofue 	 	& 30  		& $1.8$		&$6.80$  		& $1.68$   	 &4.76	 &6.52&23\\
 Xue 		&60  		&$1.5$		&$ 5.00$ 		&   $1.52$	 &5.76	&11.45&24\\
 Klypin,  A	&15  		&$0.8$		&$4.00$ 		&  $1.70$    	&3.87	&1.76&20\\
 Klypin,  B  	&15  		&$1.0$		&$5.00$		 &   $1.70$   	&4.53	&2.65&20\\
\hline 
\end{tabular}
\\ 
 \caption{ Exponent $b$ for the LCM free parameter $\alpha =  \left(\rho_{mw}/\rho_{gal}\right)^b $,  solved for by 
   power law fits to the distributions in Fig.~\ref{fig:alphaDistrib1}. 
 Integrated mass for the bulge $M_{bulge}$ and disk $M_{disk}$  are in units of  $10^{10}M_{\odot}$),  and   exponential scale lengths $r_e$  are in   kpc. $R_{last}$ is the limit of the reported Milky Way luminous mass model in   kpc. }
 \end{minipage}
  \end{table*}

 \subsection{Universal Rotation Curve trends}
  
\begin{figure*}
  \thisfloatpagestyle{empty}
 \centering
 \subfloat[][NGC 5055,   $\xi_c$]{\includegraphics[width=0.233\textwidth]{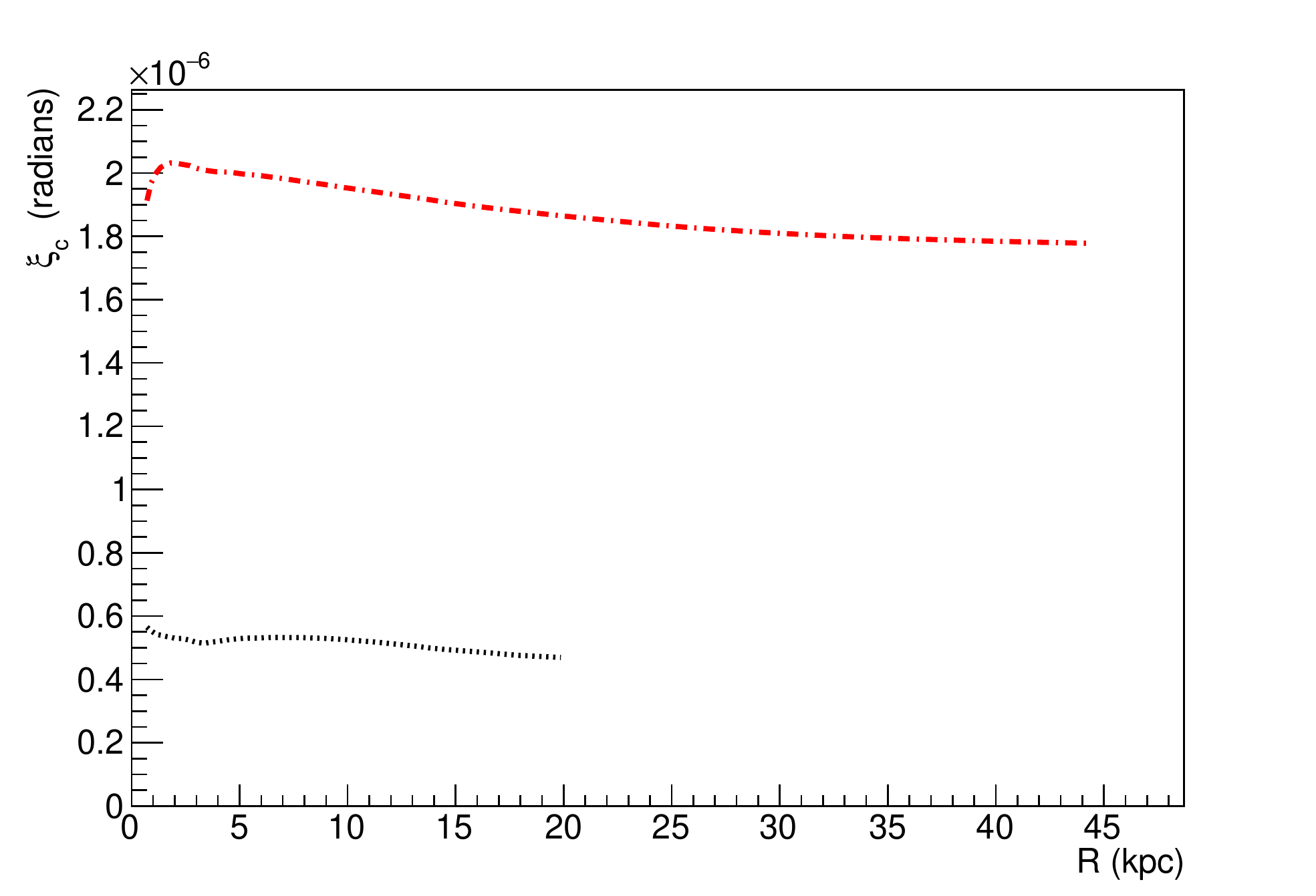}}
   \subfloat[][ NGC 7814,  $\xi_c$]{\includegraphics[width=0.23\textwidth]{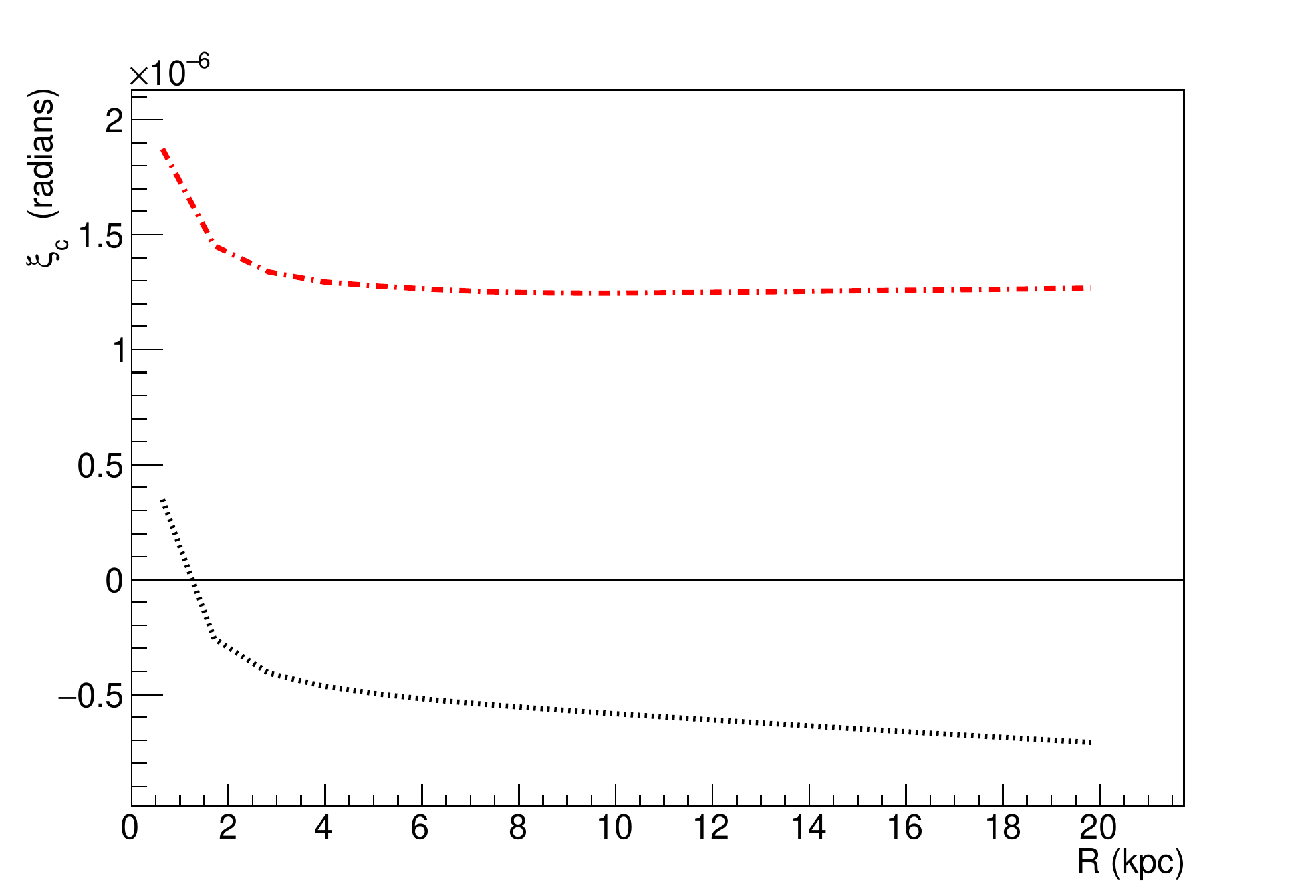}}
 \subfloat[][ NGC 7331,  $\xi_c$ ]{\includegraphics[width=0.23\textwidth]{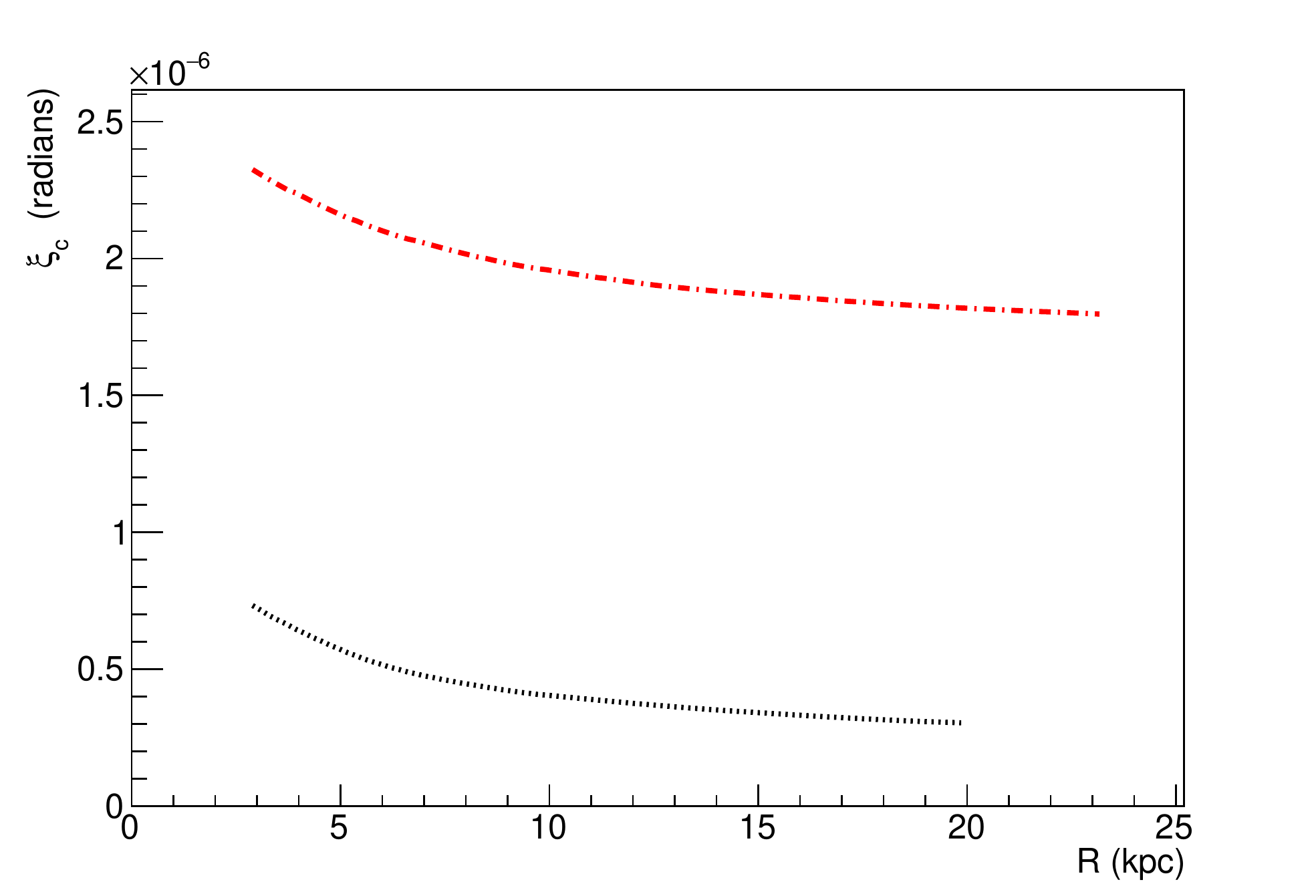}}    
 \subfloat[][ NGC 2903, $\xi_c$]{\includegraphics[width=0.23\textwidth]{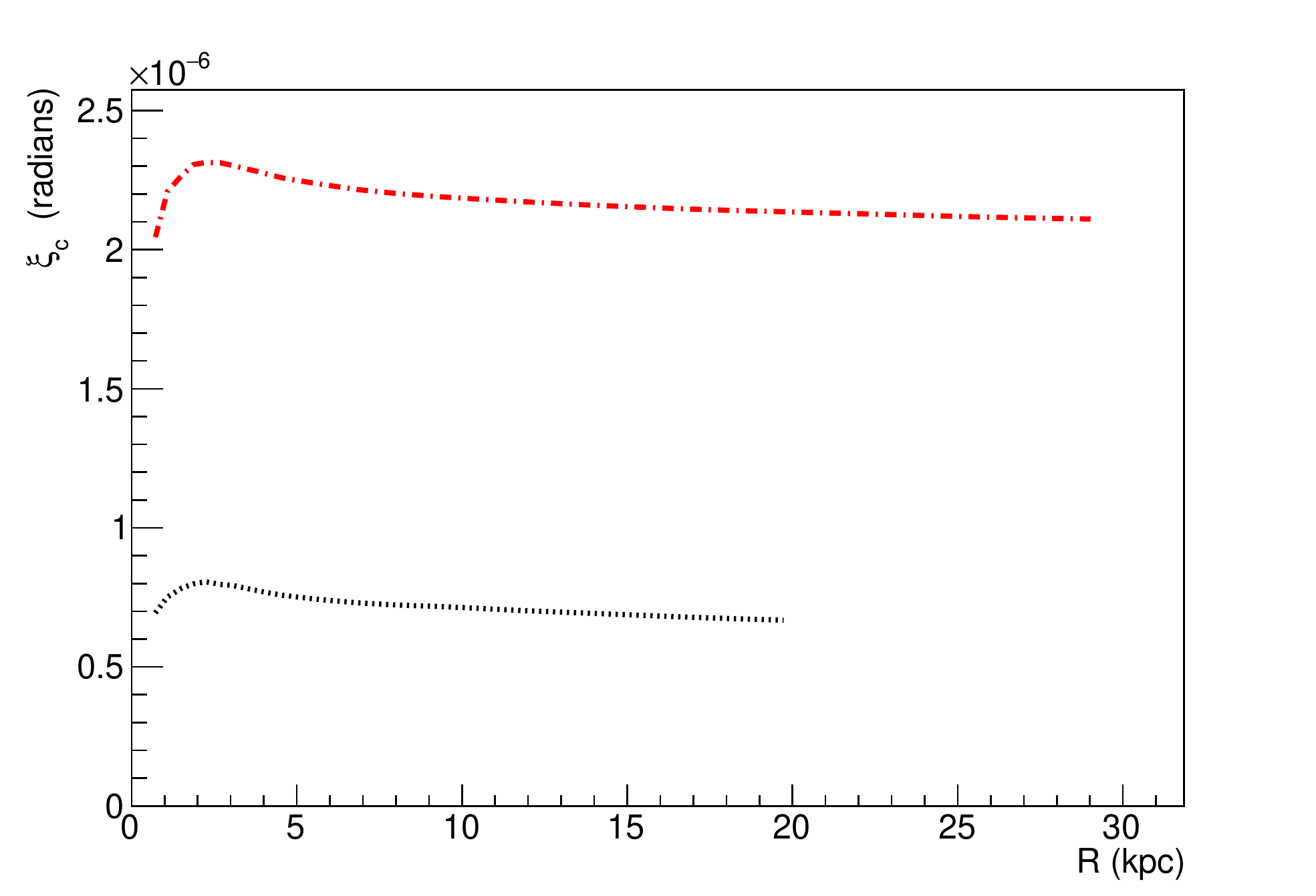}}
  \\
\vspace{0.5cm}
 \subfloat[][NGC 5055 ]{\includegraphics[width=0.23\textwidth]{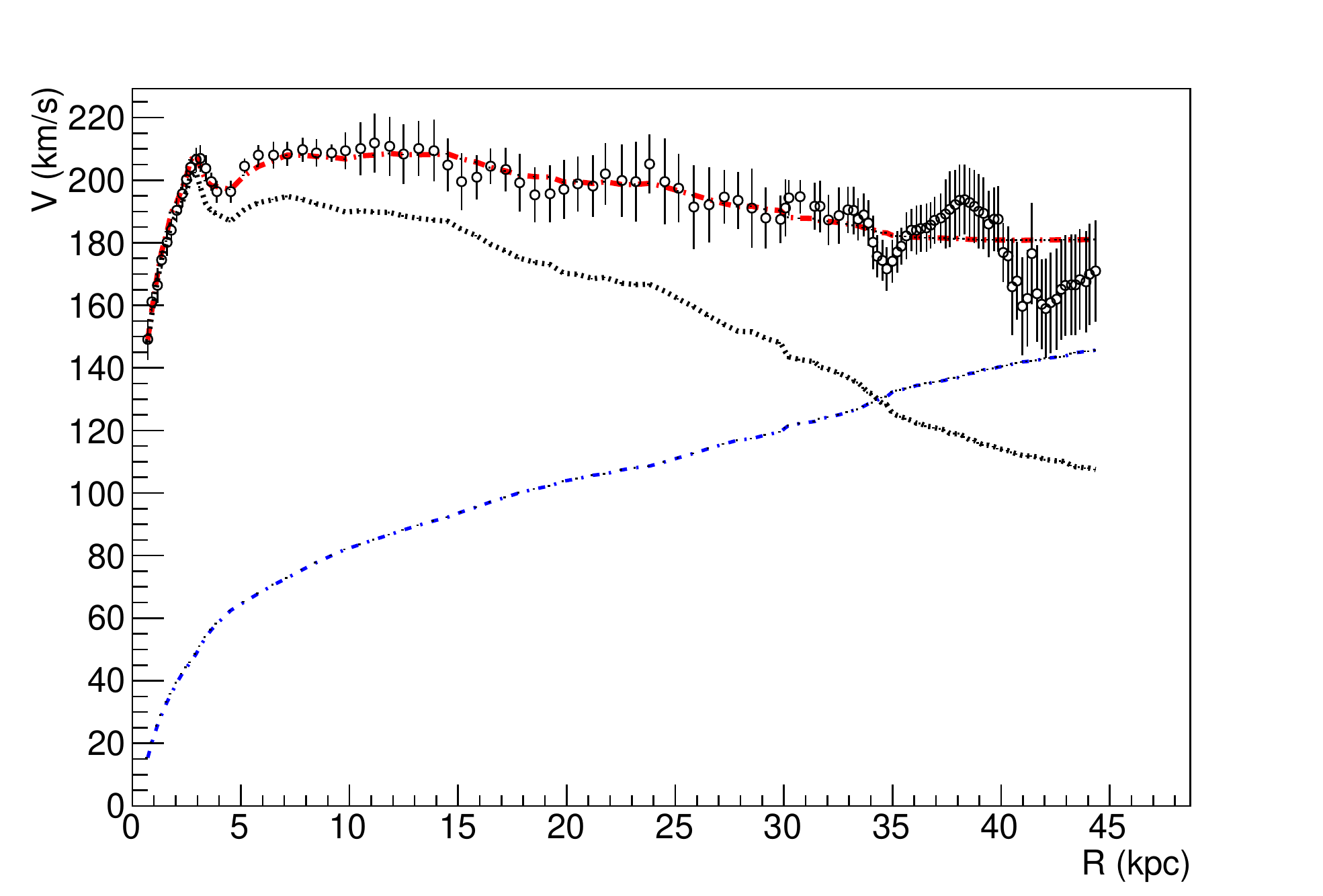}} 
  \subfloat[][ NGC 7814   ]{\includegraphics[width=0.23\textwidth]{n7814LCM.pdf}}  
   \subfloat[][ NGC 7331 ]{\includegraphics[width=0.23\textwidth]{n7331LCM.pdf}}  
 \subfloat[][ NGC 2903 ]{\includegraphics[width=0.23\textwidth]{n2903LCM}} 
  \\
\vspace{0.5cm}
  \subfloat[][NGC 3198,  $\xi_c$]{\includegraphics[width=0.23\textwidth]{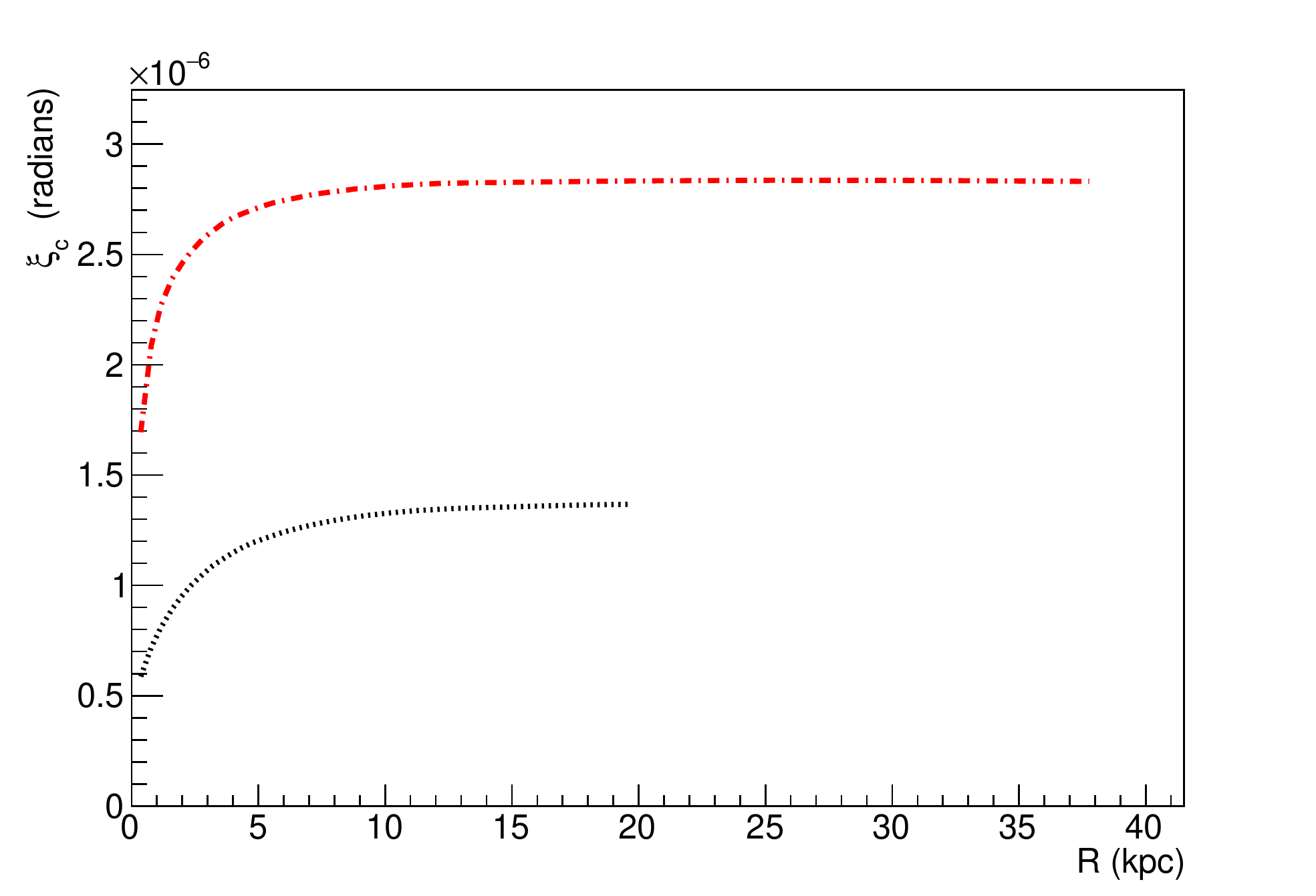}} 
 \subfloat[][ UGC 7524 ,   $\xi_c$  ]{\includegraphics[width=0.23\textwidth]{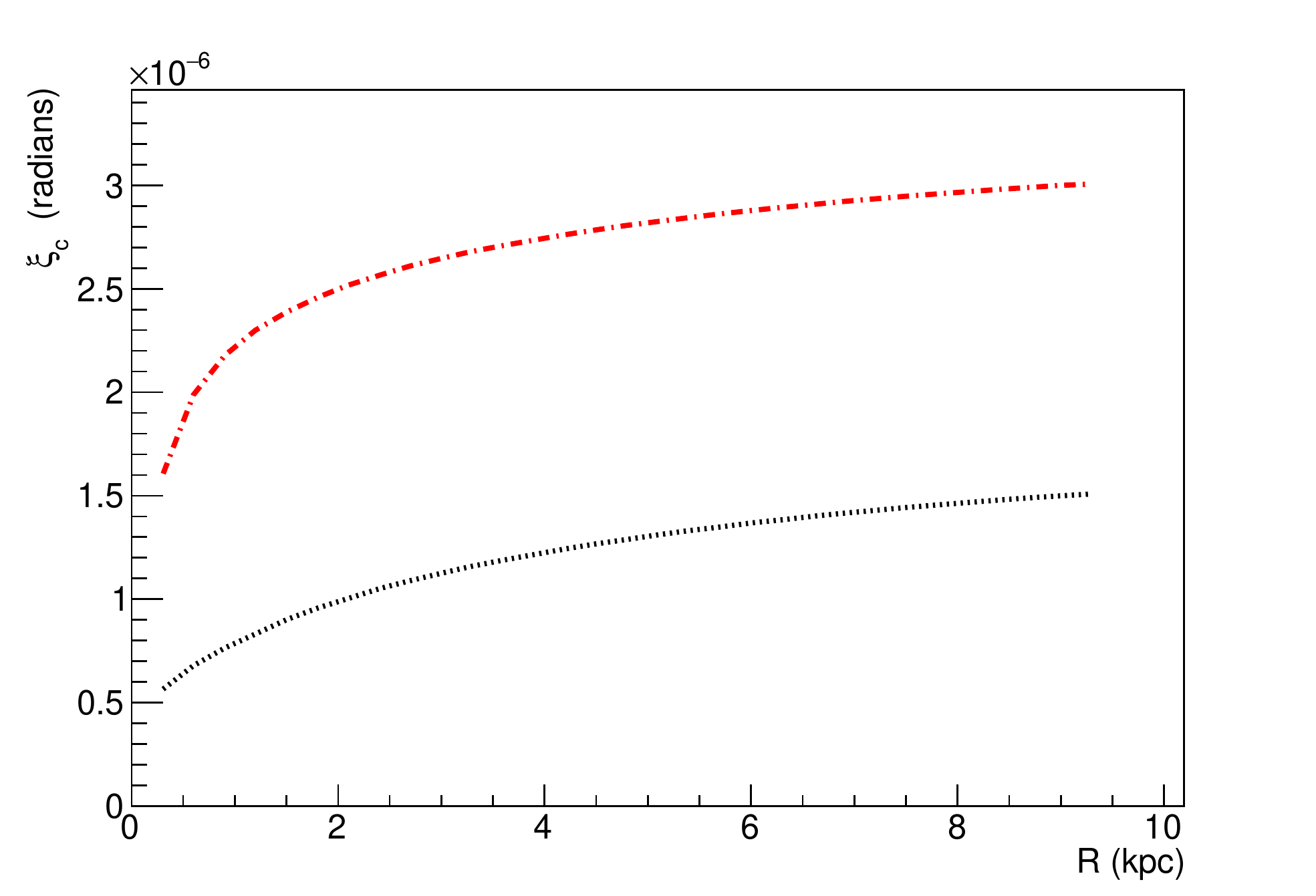}}  
 \subfloat[][ NGC 925,   $\xi_c$ ]{\includegraphics[width=0.23\textwidth]{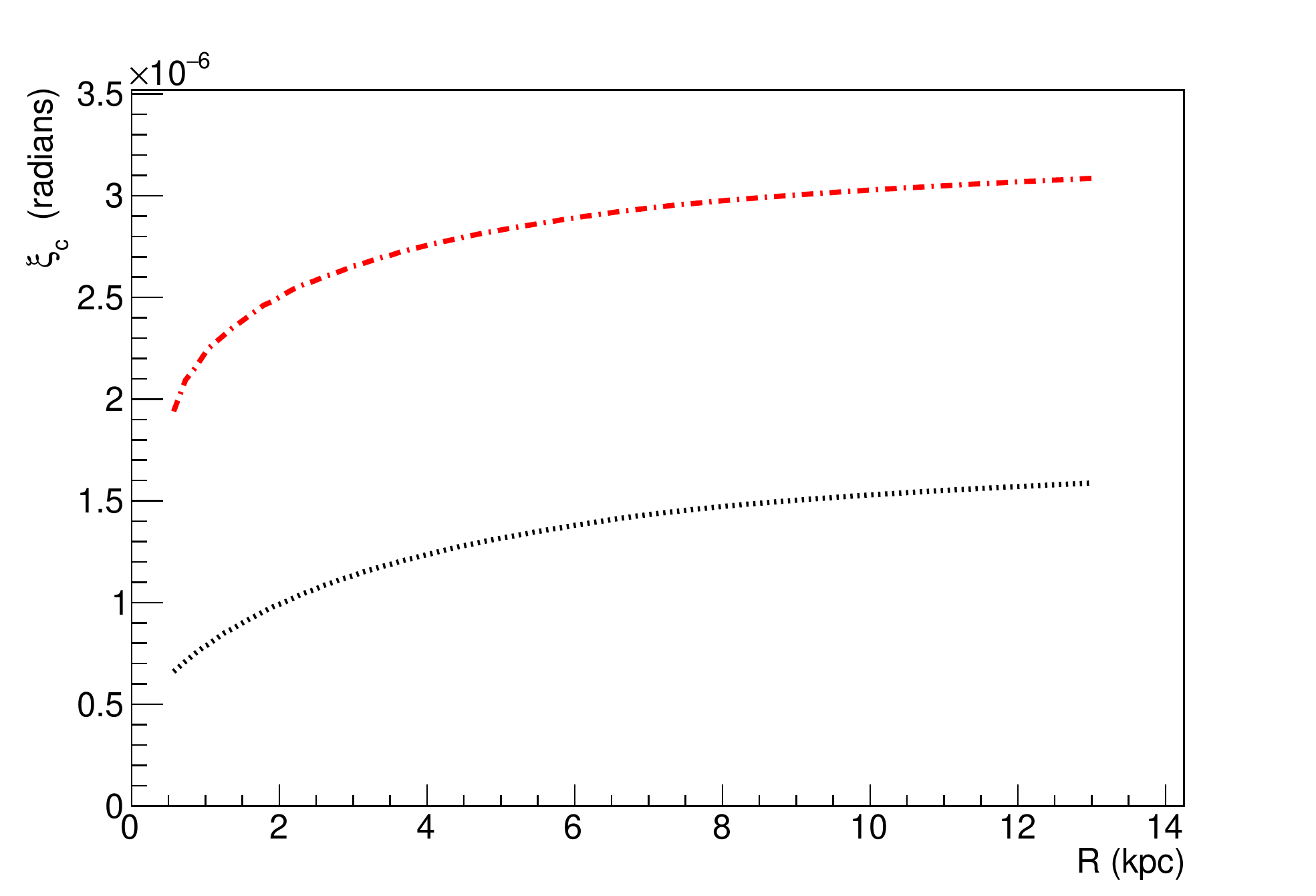}} 
   \\
\vspace{0.5cm}
 \subfloat[][NGC 3198 ]{\includegraphics[width=0.23\textwidth]{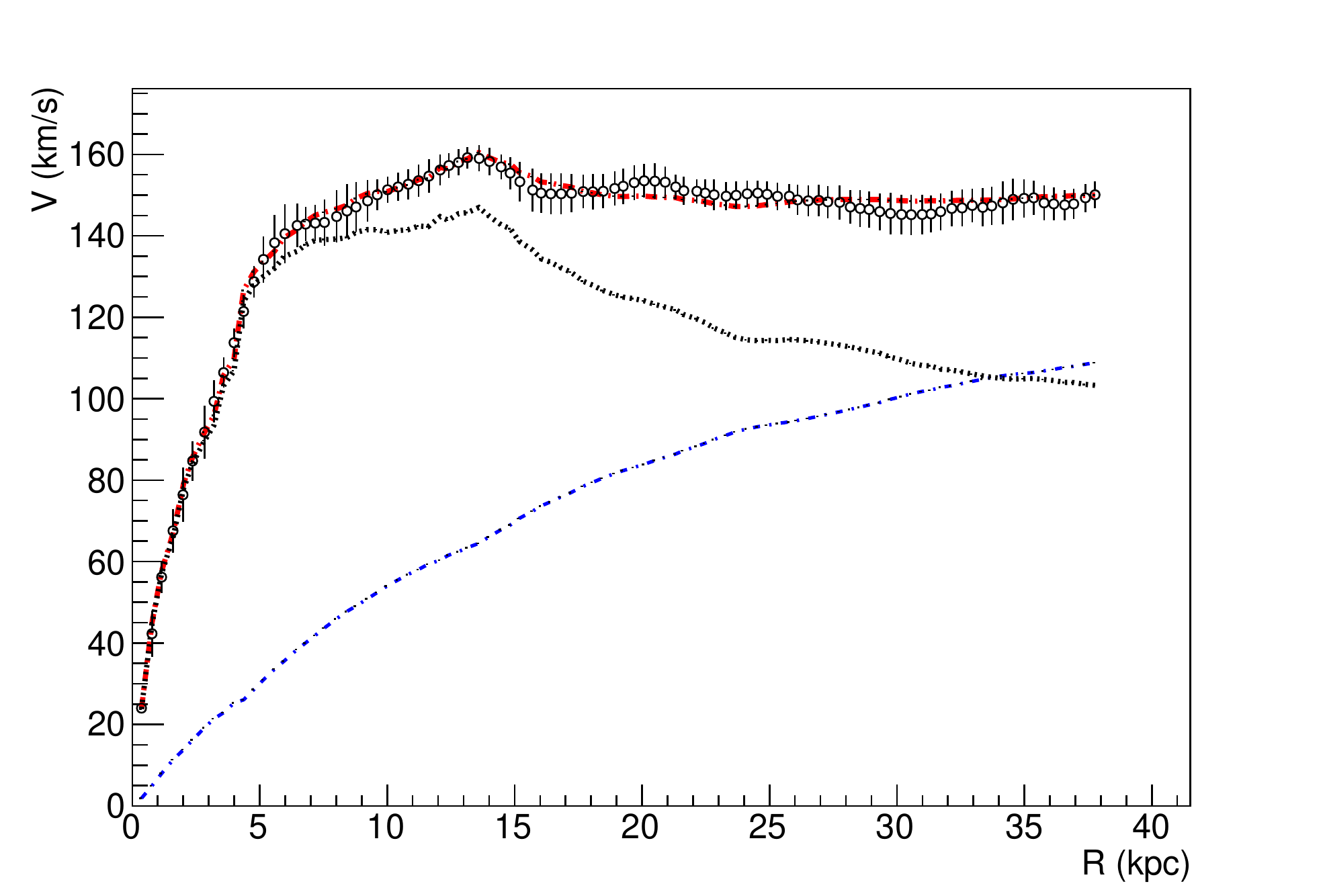}}
 \subfloat[][ UGC 7524    ]{\includegraphics[width=0.23\textwidth]{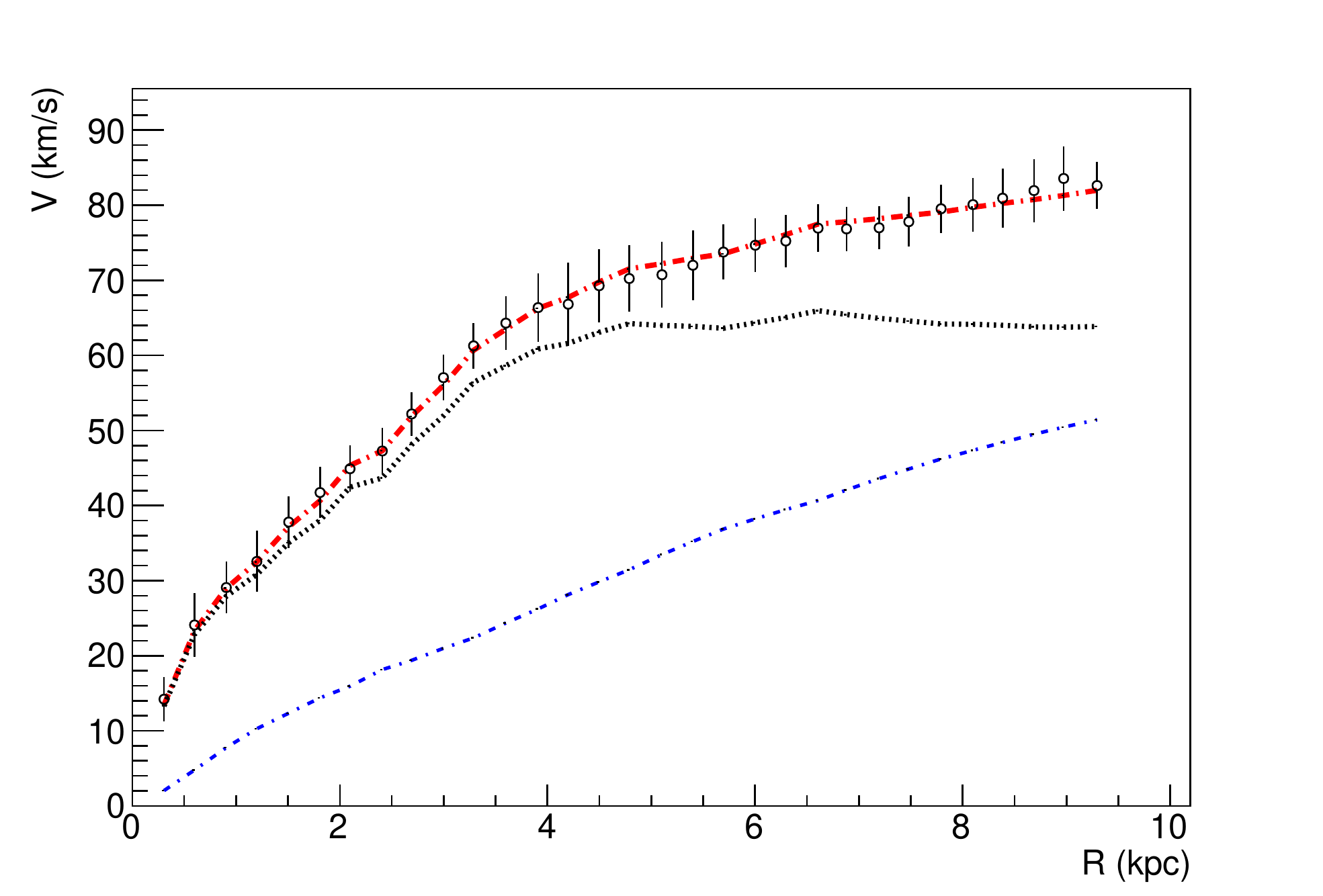}} 
 \subfloat[][ NGC 925   ]{\includegraphics[width=0.23\textwidth]{n925LCM}} 
 \caption{Universal Rotation Curve comparison:   graphs of  the Galaxy psuedo-rapidity angle  (radians) as a function of radius (kpc) in figures $(a)$-$(d)$ and $(i)$-$(k)$,  red dotted lines are for the MW choice from   \citet{Sofue} and \citet{Xue},  and black dotted lines   from \cite{Klypin}  Model B.  Rotation curves (km/s vs. kpc) are presented under corresponding galaxy $\xi_c$ pictures, and lines  and 
galaxy references are as in Fig.~\ref{galaxiesSmallest}.   }
\label{fig:results2} 
\end{figure*}

 The galaxy-to-galaxy pseudo rapidity angle $\xi_c$  (Eq.~\ref{eq:11})  gives a   simple explanation for the Universal Rotation Curve  distribution \citep{salucci}.
As can be seen in Fig.~\ref{fig:results2}, the inflection of the reported rotation curves at high radii  is   mirrored by  the  $\xi_c (r)$ function    in the galaxy disk.   The term, 
  
   \begin{equation}
   \xi_c (r) =\ln \left(\frac{\omega_{mw}(r)}{\omega_{gal}(r)} \right). 
   \label{bizness}
  \end{equation}
 is a map of the galaxy being observed onto our own Milkky Way, one-to-one in radius.  
 
In terms of  the   gravitational redshifts,  
     \begin{equation}
   \xi_c =\ln \left(\frac{\sqrt{-g_{tt, mw}(r)}}{\sqrt{-g_{tt, gal}(r)}} \right) =
   \frac{1}{2}  \ln \left(\frac{ g_{tt, mw}(r) }{ g_{tt, gal}(r)}  \right). 
   \label{bizness}
  \end{equation}
  In current analysis of the  sample,   all galaxy pseudo-rapidity angles  tend to a   constant,
 non-zero value   at high radius, on the order of $10^{-6}$, and appear to act as envelope functions to the rotation curves observed.


 \section{Conclusions}
\label{sec:conclusion}
In this paper we show how the LCM fitting formula can reconcile  observations of Doppler-shifted spectra and photometry  by the inclusion of relative galaxy   curvatures.  Relative curvature imposes the importance of the choice of a Milky Way luminous mass model, and we report our findings in this paper as to what we consider the most physical Milky Way model.   With the  unprecedented observational resolution 
of the Large Event Horizon telescope  this portion of the LCM prediction can be falsified, and could further provide a constraint to our understanding of our  Milky Way.    
The fitting formula presented in this paper is a first approach to a zero-parameter model in galaxy rotation curve modeling.  We have shown the  free parameter can be reduced to a constant value for a given choice of a  Milky Way luminous mass profile. 

Further, we have also shown that the LCM fitting formula can provide an alternate  explanation to the Universal Rotation Curve groupings   of \cite{salucci}, based only on estimates of luminous mass. 
 As can be seen in Fig.~\ref{galaxiesSmallest}, the luminous profiles used to fit the rotation curves are well within physical bounds implied by photometry.

    Although this is an attempt to explain rotation curves in a zero-parameter setting, any alternative gravitational model must be able to account for other phenomena currently attributed to cold dark matter.  Thus, in future work,  the LCM must   address questions of  weak lensing, galaxy mergers, clusters of galaxies and early structure formation.  However, none of these problems can be investigated until there is an agreed upon  estimate for the baryon distribution in the Local Group of galaxies. 
An equally interesting future question is how to phrase an LCM-style conjecture for  the rotation curve of the Milky Way itself.    Since we are within the emitter frame  it will involve a map  where the observer's frame is embedded in   the emitter's global frame.  The phenomenolgical implications of this question could provide testing much later in the future, since this prescription could be made for any emitter-receiver galaxy pair, and hence provide predictions for how rotation curves could be viewed from any arbitrary frame.

\subsection{Acknowledgments}
 
The authors would like to thank  V.\,P.\,  Nair, P.\, Schechter, R.\,A.\,M.\, Walterbos,   T.\, Boyer, P.\, Fisher,    R.\,Ott,  K.\,  Chng,   and S.\,  Rubin.   
 J.\,A. Formaggio and N.\,A. Oblath are  supported by the United States Department of Energy under Grant No. DE-FG02-06ER- 41420. 
 \bibliography{LCM}{}
\bibliographystyle{natbib}

\end{document}